\documentclass{aa}
\usepackage{graphicx}
\usepackage[varg]{txfonts}
\begin{document}
\newcommand{\be}{\begin{equation}}
\newcommand{\ee}{\end{equation}}

\title{Low redshift constraints on energy-momentum-powered gravity models}

\author{M. C. F. Faria\inst{1,2}
\and
C.J.A.P. Martins\inst{1,3}
\and
F. Chiti\inst{4}
\and
B. S. A. Silva\inst{5}}
\institute{Centro de Astrof\'{\i}sica da Universidade do Porto, Rua das Estrelas, 4150-762 Porto, Portugal\\
\email{Carlos.Martins@astro.up.pt}
\and
Faculdade de Ci\^encias, Universidade do Porto, Rua do Campo Alegre, 4150-007 Porto, Portugal\\
\email{up201604877@fc.up.pt}
\and
Instituto de Astrof\'{\i}sica e Ci\^encias do Espa\c co, CAUP, Rua das Estrelas, 4150-762 Porto, Portugal
\and
Liceo A. M. Enriques Agnotetti, Via Attilio Ragionieri 47, 50019 Sesto Fiorentino, Italy
\and
Instituto Superior T\'ecnico, Universidade de Lisboa, Avenida Rovisco Pais, 1049-001 Lisboa, Portugal}
\date{Submitted \today}

\abstract
{There has been recent interest in the cosmological consequences of energy-momentum-powered gravity models, in which the matter side of Einstein's equations is modified by the addition of a term proportional to some power, $n$, of the energy-momentum tensor, in addition to the canonical linear term. In this work we treat these models as phenomenological extensions of the standard $\Lambda$CDM, containing both matter and a cosmological constant. We also quantitatively constrain the additional model parameters using low redshift background cosmology data that are specifically from Type Ia supernovas and Hubble parameter measurements. We start by studying specific cases of these models with fixed values of $n,$ which lead to an analytic expression for the Friedmann equation; we discuss both their current constraints and how the models may be further constrained by future observations of Type Ia supernovas for WFIRST complemented by measurements of the redshift drift by the ELT. We then consider and constrain a more extended parameter space, allowing $n$ to be a free parameter and considering scenarios with and without a cosmological constant. These models do not solve the cosmological constant problem per se. Nonetheless these models can phenomenologically lead to a recent accelerating universe without a cosmological constant at the cost of having a preferred matter density of around $\Omega_M\sim0.4$ instead of the usual $\Omega_M\sim0.3$. Finally we also briefly constrain scenarios without a cosmological constant, where the single component has a constant equation of state which needs not be that of matter; we provide an illustrative comparison of this model with a more standard dynamical dark energy model with a constant equation of state.}


\titlerunning{Low redshift constraints on energy-momentum-powered gravity models}
\authorrunning{Faria \& et al.}
\maketitle

%%%%%%%%%%%%%%%%%%%%%%%%%%%%%%%%%%%%%%%%%%%%%%%%%%%%%%%%%%%%%%%%%%%%%%%%%%
\section{Introduction}
\label{introd}

The search for the physical mechanism behind the observed recent acceleration of the universe is among the most compelling tasks of modern physics and cosmology. While a cosmological constant is the simplest known possibility, its well-known  problems of fine-tuning motivate the search for theoretical or phenomenological  alternatives, which may be tested by rapidly improving astrophysical and cosmological data.

Recently \citet{Roshan} have studied a model of the so-called energy-momentum-squared gravity, where the matter part of Einstein's equations is modified by the addition of a term proportional to $T^2\equiv T_{\alpha\beta}T^{\alpha\beta}$ and $T_{\alpha\beta}$ is the energy-momentum tensor. The authors showed that depending on the choice of the proportionality factor the model may have a cosmologically interesting early-time behavior, such as a bounce, which would avoid the initial singularity. Subsequent works by \citet{Board} and \citet{Akarsu} have extended this to the more generic form $(T^2)^n$, dubbed energy-momentum-powered gravity. Qualitatively, we may think of these models as extensions of general relativity with a nonlinear matter Lagrangian. As such they are somewhat different from the usual dynamical dark energy or modified gravity models: in the former class of models  further dynamical degrees of freedom to the Lagrangian (often in the form of scalar fields) are added, while in the latter the gravitational part of the Lagrangian is  changed.

The initial studies of the cosmological implications of these models focused on their early-time behavior, which indeed has several interesting features. In this work we take a complementary approach and focus on the low redshift behavior of these models; we study them in the context of the recent acceleration of the universe using some of the latest low redshift data from the Pantheon Type Ia supernova compilation by \citet{Riess} and the compilation of 38 Hubble parameter measurements by \citet{Farooq} to constrain these models. These are our baseline datasets, but we additionally discuss how the constraints change if we use alternative datasets or complement the baseline datasets with external priors.

In practical phenomenological terms, we may think of these models as extensions to the canonical $\Lambda$CDM, in which case the model still has a cosmological constant but the nonlinear matter Lagrangian leads to additional terms in Einstein's equations, and cosmological observations can therefore constrain the corresponding additional model parameters in these terms. Typically there are two such additional parameters: the power $n$ of the nonlinear part of the Lagrangian and a further parameter (to be defined below) quantifying the contribution of this term to the energy budget of the universe. We study this scenario, starting with few cases with fixed values of $n$ where we can find analytic solutions, since in these cases the interpretation of the extra terms appearing in the Friedmann equation are physically clear. We constrain these models using current data, and also discuss how they can be further constrained by future observations of Type Ia supernovas from the proposed WFIRST satellite \citep{WFIRST} together with measurements of the redshift drift by the Extremely Large Telescope (ELT) \citep{Liske,HIRES}. Later in the work we also study constraints on the whole class of these models.

We may also ask whether a suitably chosen nonlinear Lagrangian can reproduce the recent (low redshift) acceleration of the universe in a model which at low redshift only contains matter (plus a subdominant amount of radiation) but no true cosmological constant. In principle such a scenario is conceivable and has been qualitatively discussed in the original work \citep{Roshan}. It is also somewhat closer in spirit to the usual modified gravity models with the caveat that, as previously mentioned, in the latter models the modification occurs in the gravitational part of the Lagrangian and not in the matter part. We show that these models have a phenomenological limit where this is possible, and we also provide constraints on this scenario.

The structure of the rest of this work is as follows. We start in Sect. \ref{mods} with a brief but generic overview of these models and then discuss their low redshift limit and specifically the three choices of $n$ that lead to relatively simple analytic solutions. We then constrain these models, treated as extensions of $\Lambda$CDM, in Sect. \ref{cnstr} using current low redshift data, starting with our baseline datasets of Pantheon supernova and Hubble parameter measurements. We also study how these constraints change with different datasets; that is, specifically replacing Union2.1 supernovas with Pantheon supernovas, or replacing the Hubble parameter measurements by a prior on the matter density. In Sect. \ref{frcst} we briefly study how constraints on these models will be improved by future astrophysical facilities. In Sect. \ref{genn} we consider and constrain the extended parameter space, where $n$ is a free parameter. In this section, we again consider the $\Lambda$CDM extension scenario, but we also discuss the extent to which these models can phenomenologically lead to a recent accelerating universe without a true cosmological constant. For this latter case, we also consider the case in which the equation of state of the single component, which we assume to be constant, need not be that of matter. Finally, we present some conclusions in Sect. \ref{concl}. We also provide an illustrative comparison of this model with a more standard dynamical dark energy model with a constant equation of state in a brief Appendix.

%%%%%%%%%%%%%%%%%%%%%%%%%%%%%%%%%%%%%%%%%%%%%%%%%%%%%%%%%%%%%%%%%%%%%%%%%%
\section{Generic Einstein equations and simple solutions}
\label{mods}

The general action for these models has the form \citep{Roshan,Board}
\be
S=\frac{1}{2\kappa}\int\left[R+\eta (T^2)^n-2\Lambda\right]d^4x + S_{matter}\,,
\ee
where $\kappa=8\pi G$, $\eta$ is a constant quantifying the contribution of the $T$-dependent term, and $n=1$ is the particular case discussed in \citet{Roshan}, which also suggests that $\eta<0$ leads to an early-universe bounce and satisfactory late-time behavior. Conversely, these authors suggested that for $\eta>0$ this behavior is unsatisfactory since there is no stable late-time acceleration; however, we note that this division is less straightforward for generic $n\neq1$. In passing we also mention that at a phenomenological level a further possibility would be to have a term proportional to $(T_a^a)^2$, as studied by \citet{Poplawski}; we leave the study of this alternative for future work. 

In a flat Friedmann-Lemaitre-Robertson-Walker universe and assuming a perfect fluid, the generalized Friedmann and Raychaudhuri equations and the corresponding continuity equation can be written
as\be
3\left(\frac{\dot a}{a}\right)^2=\Lambda+\kappa\rho+\eta(\rho^2+3p^2)^{n-1}\left[(n-\frac{1}{2})(\rho^2+3p^2)+4np\rho\right]
\ee
\be
6\frac{\ddot a}{a}=2\Lambda-\kappa(\rho+3p)-\eta(\rho^2+3p^2)^{n-1}\left[(n+1)(\rho^2+3p^2)+4np\rho\right]
\ee
\be
{\dot\rho}=-3\frac{\dot a}{a}(\rho+p)\frac{\kappa\rho+n\eta\rho(\rho+3p)(\rho^2+3p^2)^{n-1}}{\kappa\rho+2n\eta(\rho^2+3p^2)^{n-1}\left[\left(n-\frac{1}{2}\right)(\rho^2+3p^2)+4np\rho\right]}.
\ee
As usual, the Bianchi identity implies that only two of these equations are independent; for our purposes in the present work, the most convenient choice is to use the Friedmann and continuity equations.

In what follows we consider the low redshift limit of these models, further assuming that the universe is composed of matter and possibly also a cosmological constant. In doing so we are phenomenologically treating these models as one-parameter extensions of the canonical $\Lambda$CDM model, in which an additional $\eta$-dependent term is constrained by observations. In this case we can simplify the Einstein equations to
\be
3\left(\frac{\dot a}{a}\right)^2=\Lambda+\kappa\rho+(n-\frac{1}{2})\eta \rho^{2n} \label{maineq1}
\ee
\be
6\frac{\ddot a}{a}=2\Lambda-\kappa\rho-(n+1)\eta\rho^{2n}\,,
\ee
while the continuity equation becomes
\be
{\dot\rho}=-3\frac{\dot a}{a}\rho\frac{\kappa+n\eta\rho^{2n-1}}{\kappa+(2n-1)n\eta\rho^{2n-1}}. \label{maineq2}
\ee
We note that the Friedmann equation has some phenomenological similarities with the Cardassian models introduced by \citet{Cardassian}. Broadly speaking, inspection of the equations would suggest that $n>1/2$ may have nontrivial impacts at early times, while $n<1/2$ may be interesting at late times; we confirm these expectations in what follows.

In general these equations need to be solved numerically. However, there are three particular cases for which analytic solutions can be found (at least approximate, low redshift solutions), corresponding to the values $n=1$, $n=1/2$ and $n=0$. These have been studied, in a general mathematical context by \citet{Roshan}, \citet{Early}, and \citet{Board}, respectively. In this work we revisit these cases, describing them more systematically, emphasizing the physical context, and expressing them in a way that is directly testable with low redshift data, which we do in the next section. Later in the article we also discuss the more general case.

\subsection{The case $n=1$}
\label{mods1}

This was the case originally considered by \citet{Roshan}. In this case the continuity equation leads to the usual behavior for the matter density
\be
\rho=\rho_0a^{-3}
\ee
and we can write the Friedmann equation as a function of redshift (defined as $1+z=1/a$) as
\be
\frac{H^2(z)}{H_0^2}=\Omega_\Lambda+\Omega_M(1+z)^3+\Omega_Q(1+z)^6\,,
\ee
with the flatness assumption requiring $\Omega_\Lambda+\Omega_M+\Omega_Q=1$, so only two of these parameters are independent. For convenience we have defined
\be
\Omega_\Lambda=\frac{\Lambda}{3H_0^2}
\ee
\be
\Omega_M=\frac{\kappa\rho_0}{3H_0^2}
\ee
\be
\Omega_Q=\frac{\eta\rho_0^2}{6H_0^2}\,.
\ee
Alternatively, and for easier comparison with the other choices of $n$, we define
\be
Q_1=\frac{\eta\rho_0}{2\kappa}\,,
\ee
leading to a Friedmann equation of the form
\be
\frac{H^2(z)}{H_0^2}=\Omega_\Lambda+\Omega_M(1+z)^3+Q_1\Omega_M(1+z)^6\,.
\ee

For this choice of $n$ the effect of the nonlinear matter Lagrangian is a kination-type term behaving as $a^{-6}$. It is clear that this is tightly constrained (being problematic at high redshifts) and that by itself it cannot lead to late-time acceleration. Nevertheless it is a simple and useful toy model to test how well such terms can be constrained by low redshift data alone. We note in particular that sufficiently negative values of $\Omega_Q$ or $Q_1$ would lead to negative values of the Hubble parameter, which is certainly pathological.

\subsection{The case $n=1/2$}
\label{mods2}

This case was briefly considered by \citet{Early} in the context of studies of $f(R)$ models with which it has some similarities. In this case the continuity equation leads to
\be
\rho=\rho_0a^{-3(1+Q_{1/2})}\,,
\ee
where we defined
\be
Q_{1/2}=\frac{\eta}{2\kappa}\,,
\ee
and the Friedmann equation has the form
\be
\frac{H^2(z)}{H_0^2}=\Omega_\Lambda+\Omega_M(1+z)^{3(1+Q_{1/2})}\,,
\ee
and in the flat case we have the further condition $\Omega_\Lambda+\Omega_M=1$. In this case we have a modified evolution of the scale factor in the matter era, which again should be tightly constrained; a recent analysis in the context of other extensions of the standard cosmological model can be found in \citet{Tutusaus}.

\subsection{The case $n=0$}
\label{mods3}

This case was briefly studied in \citet{Board} and in \citet{Akarsu}. The latter did a first comparison of the model with 28 measurements of the Hubble parameter; we update this analysis with additional data.

In this case the continuity equation also leads to the standard solution
\be
\rho=\rho_0a^{-3}
\ee
and a Friedmann equation is
\be
\frac{H^2(z)}{H_0^2}=(\Omega_\Lambda-Q_0\Omega_M)+\Omega_M(1+z)^3\,, \label{enzero}
\ee
where we also defined
\be
Q_0=\frac{\eta}{\kappa\rho_0}\,.
\ee
In this case in addition to the usual cosmological constant $\Omega_\Lambda$ there is a further constant term given by $-Q_0\Omega_M$. We may therefore ask if one could do without the usual cosmological constant and still obtain an accelerating universe compatible with observations. We address and answer this question later in the article. On the other hand, the flatness constraint is $\Omega_\Lambda+(1-Q_0)\Omega_M=1$, which upon substitution leads to
\be
\frac{H^2(z)}{H_0^2}=(1-\Omega_M)+\Omega_M(1+z)^3\,; \label{enzeronoq}
\ee
in other words, this corresponds to the $\Lambda$CDM case.

%%%%%%%%%%%%%%%%%%%%%%%%%%%%%%%%%%%%%%%%%%%%%%%%%%%%%%%%%%%%%%%%%%%%%%%%%%
\section{Constraints for specific values of $n$}
\label{cnstr}

In this section we discuss constraints on the models introduced in the previous section, assuming flatness and also a non-zero cosmological constant. The models are therefore extensions of $\Lambda$CDM, to which they reduce in the appropriate limits of the model parameters $Q_n$.

\subsection{Baseline datasets}
\label{basel}

We used the compressed data from the Pantheon compilation in \citet{Riess}. We note that the values reported in the arxiv and published versions are slightly different; in what follows we used the values from the published version. The 1049 supernova measurements in the range $0<z<2.3$ are compressed into 6 correlated measurements of $E^{-1}(z)$ (where $E(z)=H(z)/H_0$) in the redshift range $0.07<z<1.5$, providing a nearly identical characterization of dark energy as the full supernova sample, thus making it an efficient compression of the raw data. Additionally we used the compilation of Hubble parameter measurements from \citet{Farooq}.

\begin{figure*}
\centering
\includegraphics[width=7.5cm]{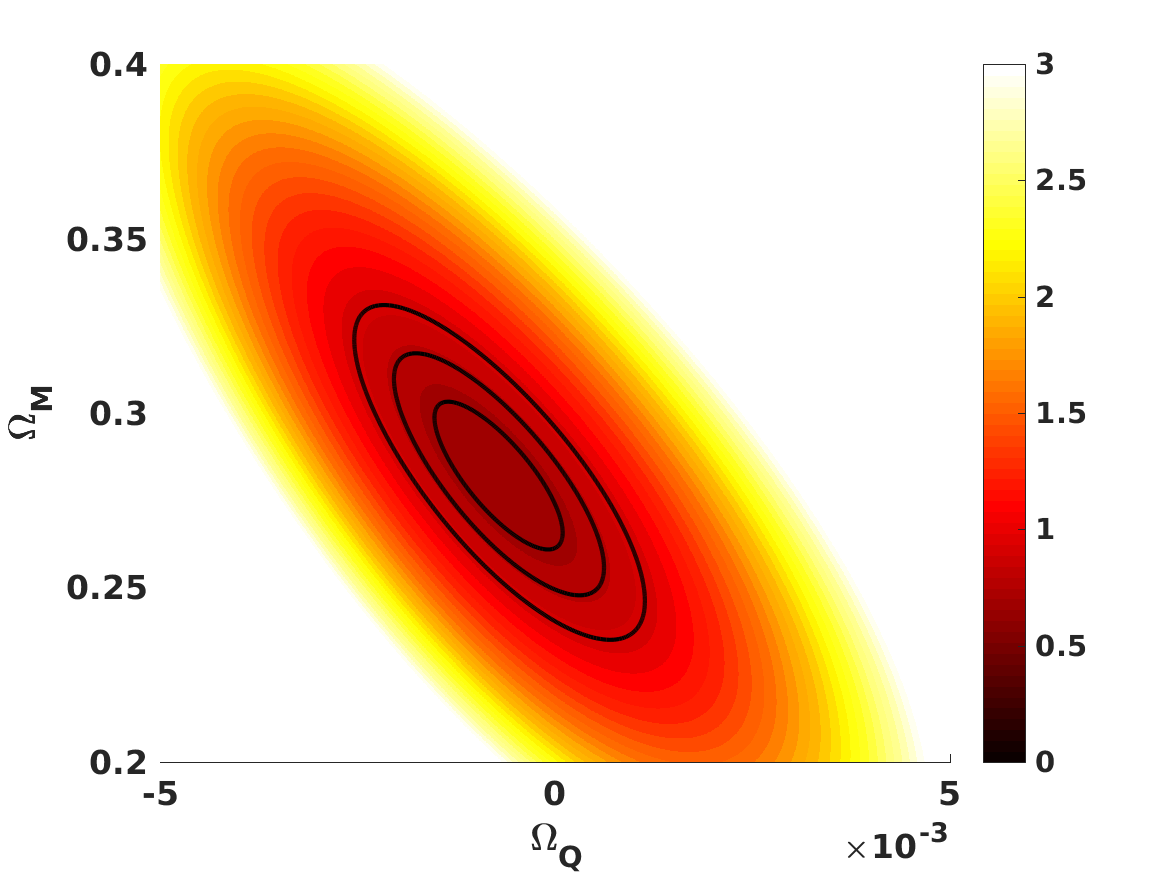}
\includegraphics[width=7.5cm]{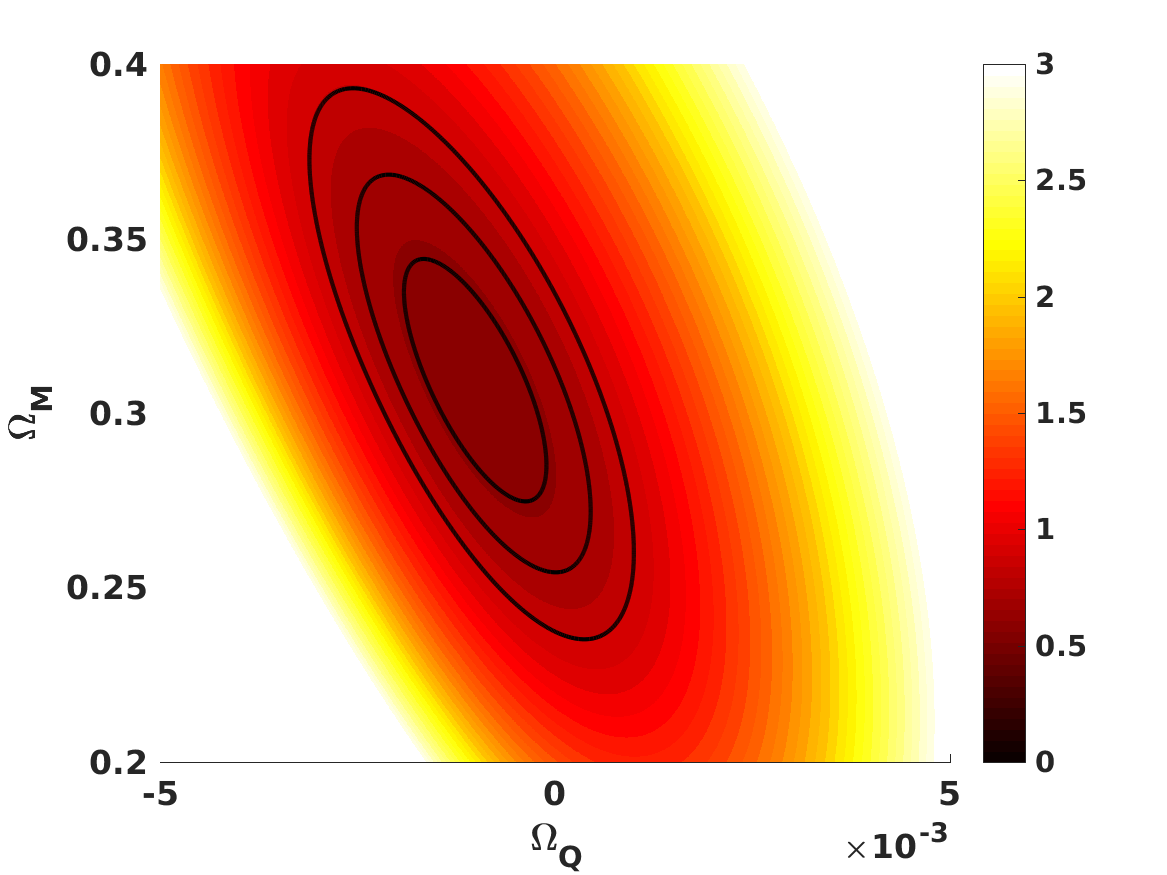}
\includegraphics[width=7.5cm]{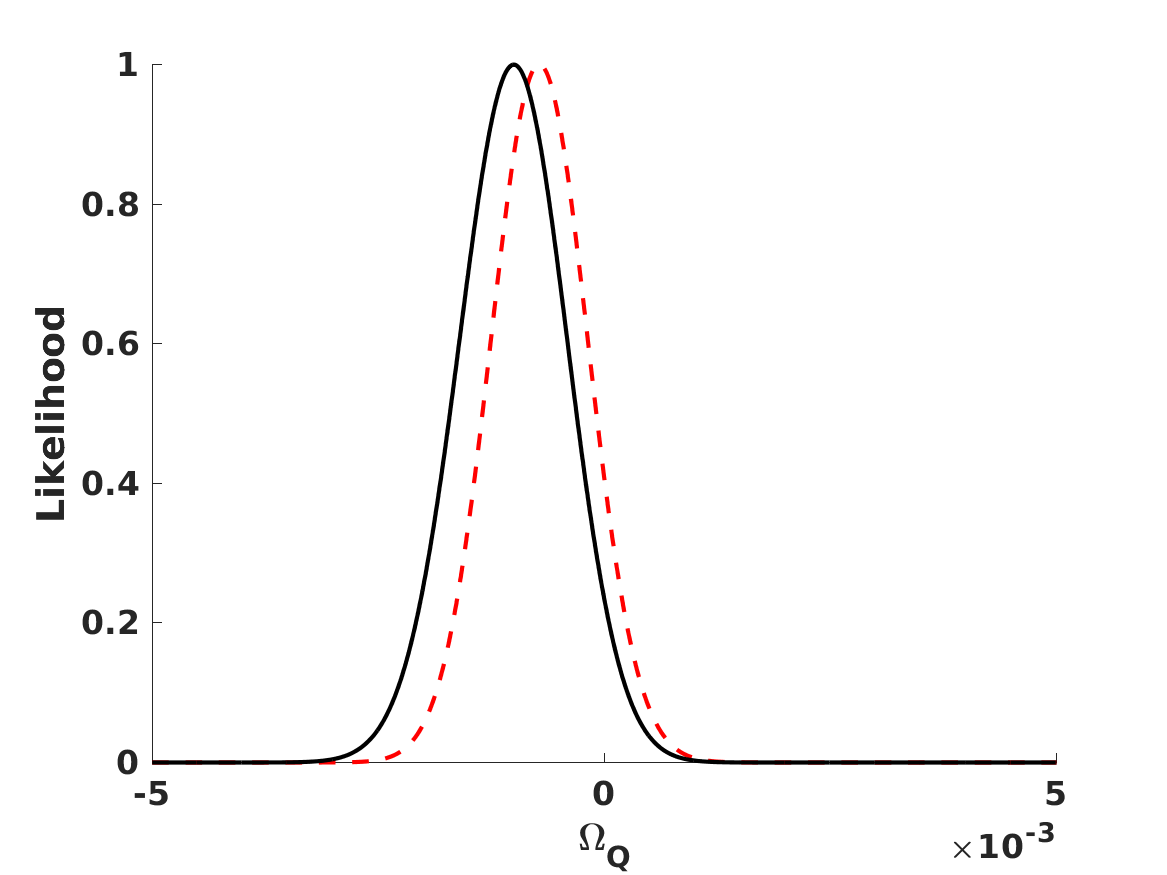}
\includegraphics[width=7.5cm]{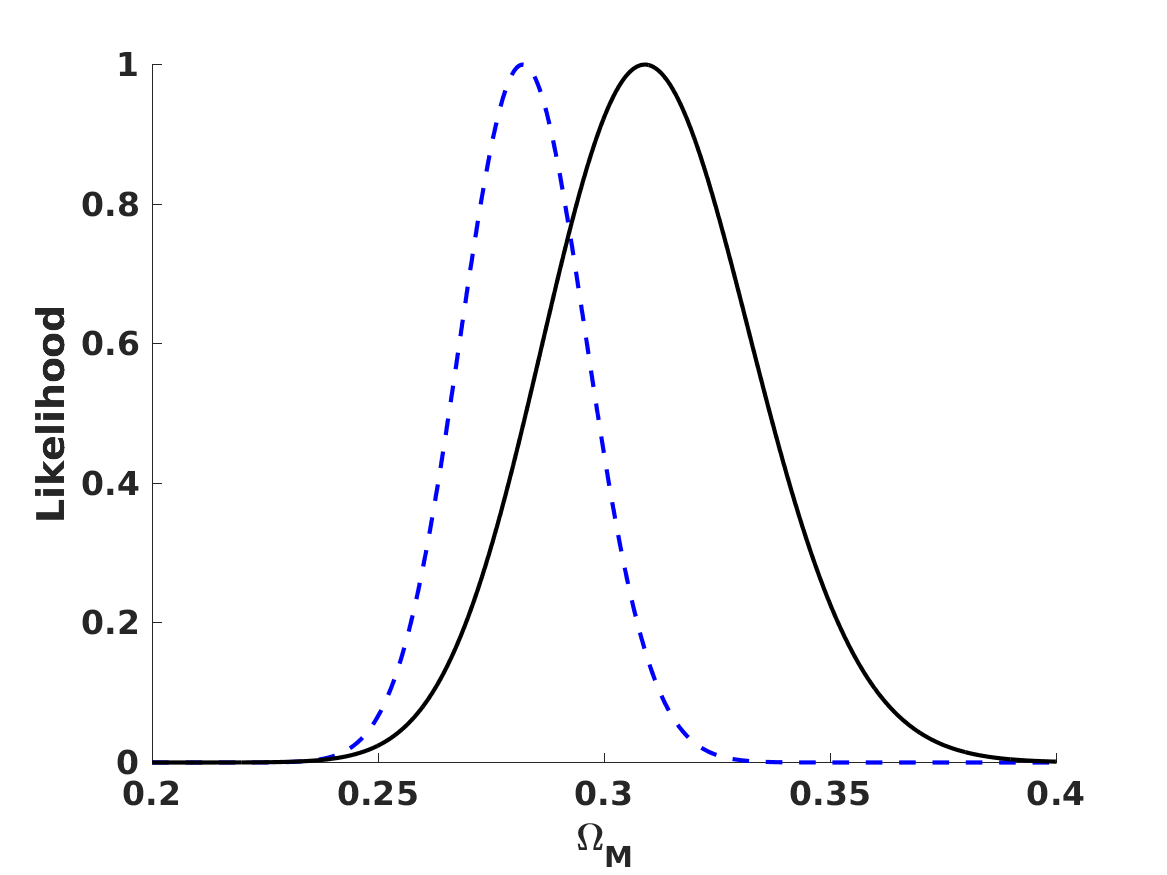}
\caption{Current constraints for the $n=1$ case, using $\Omega_Q$ as free parameter. In the top panels the black solid curves show the one, two, and three sigma confidence levels in the two-dimensional plane, while the color map depicts the reduced chi-square; the left panel is for the case of a fixed $H_0$, while in the right panel it has been analytically marginalized. The bottom panels show the one-dimensional posterior likelihoods for both parameters; the dashed lines correspond to the fixed $H_0$ and the solid lines to the marginalized case.}
\label{figure1}
\end{figure*}

\begin{figure*}
\centering
\includegraphics[width=7.5cm]{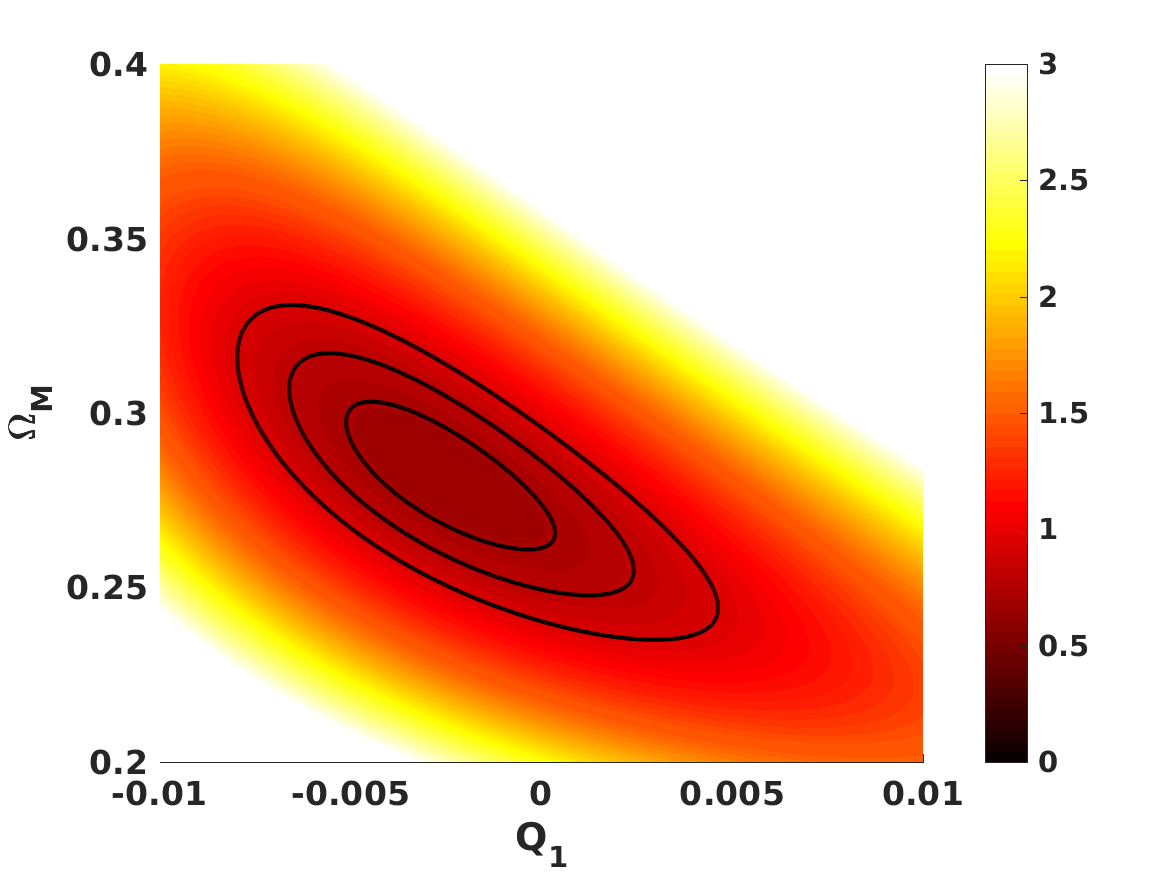}
\includegraphics[width=7.5cm]{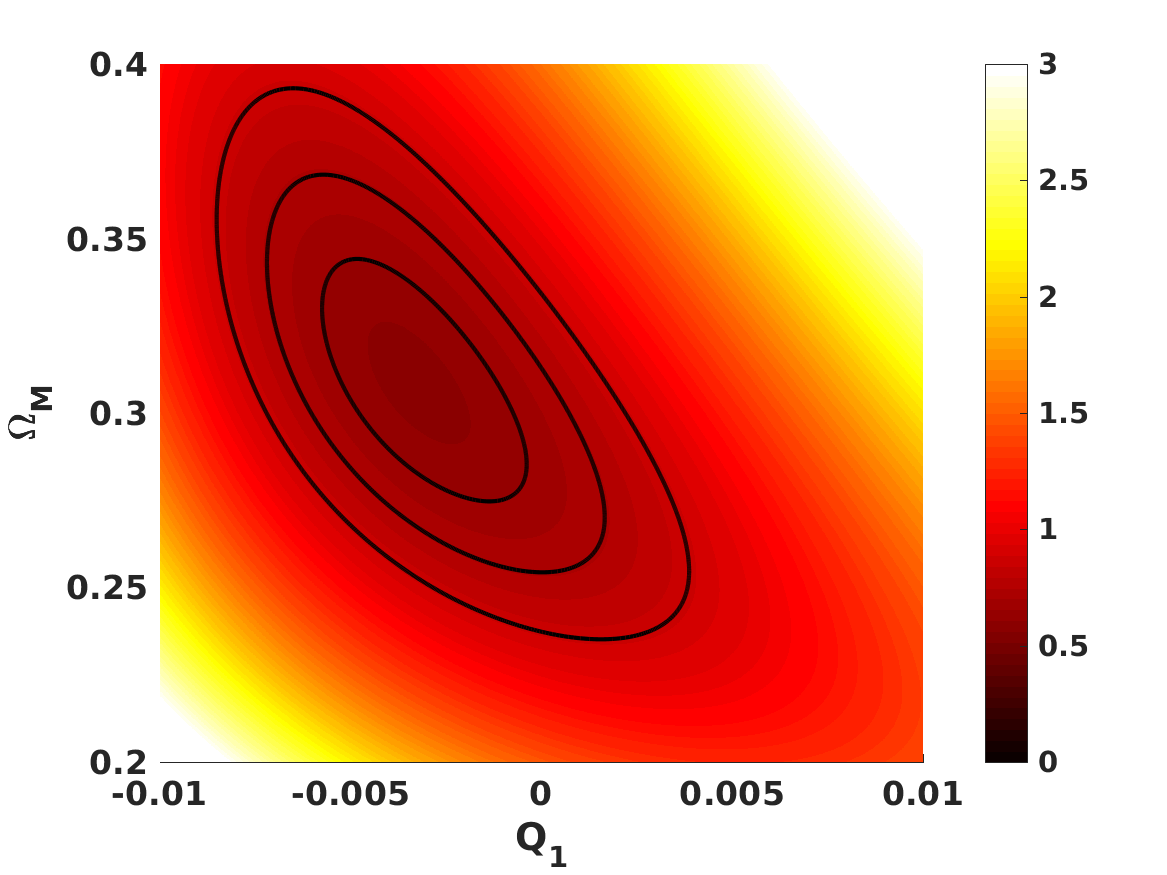}
\includegraphics[width=7.5cm]{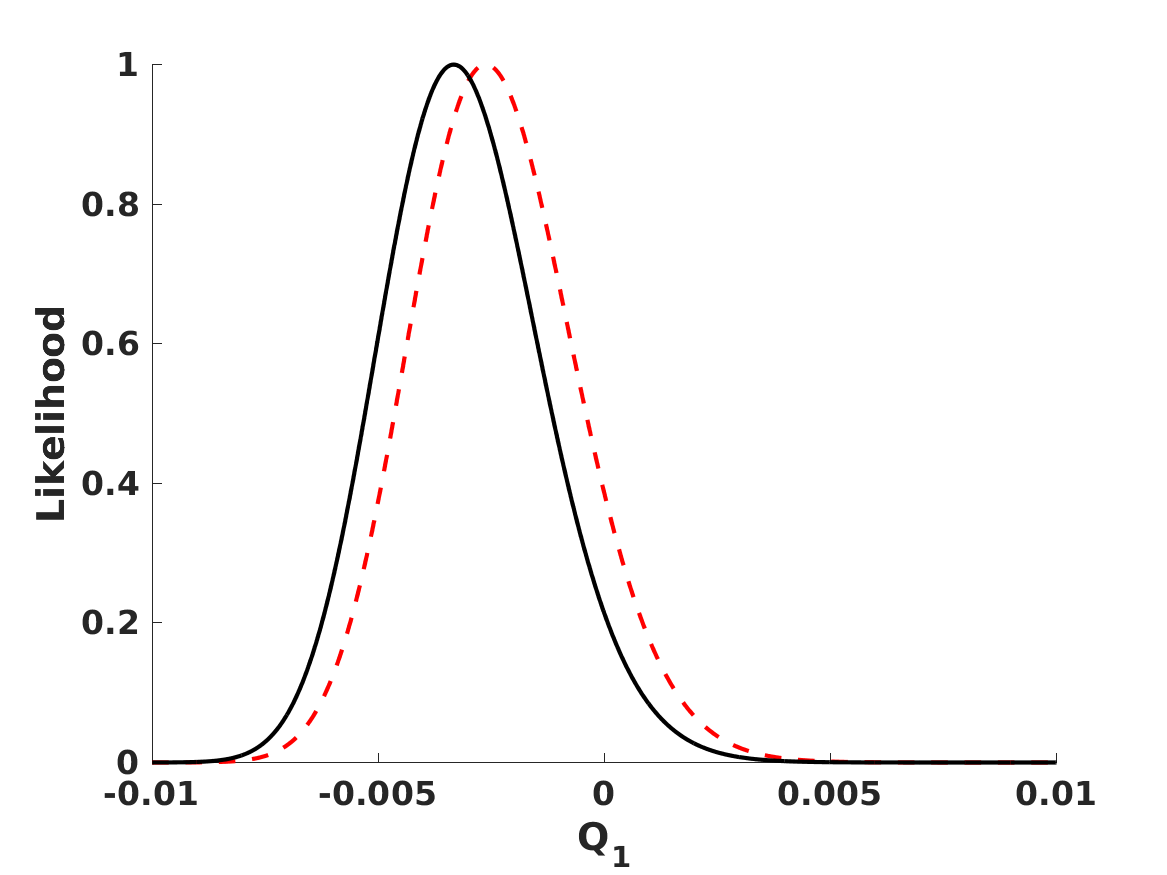}
\includegraphics[width=7.5cm]{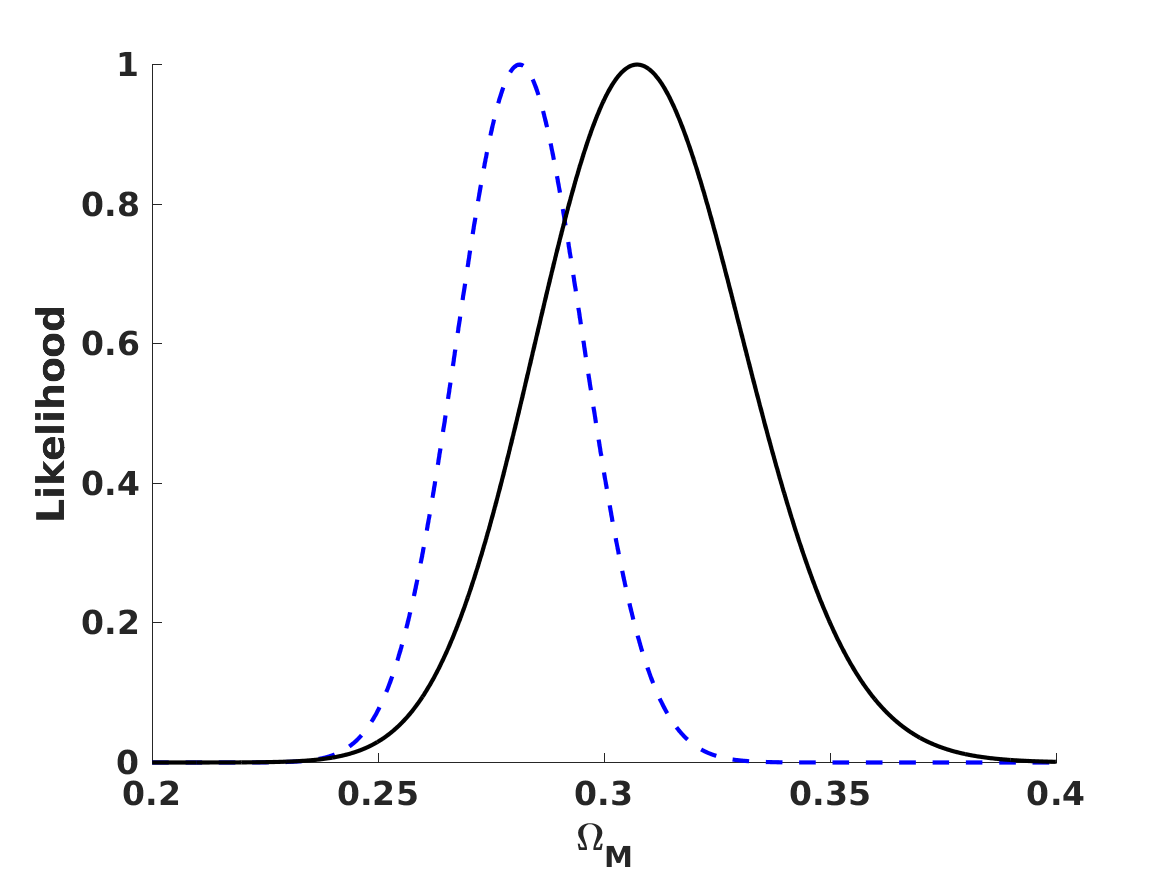}
\caption{Current constraints for the $n=1$ case, using $Q_1$ as free parameter. In the top panels the black solid curves show the one, two, and three sigma confidence levels in the two-dimensional plane, while the color map depicts the reduced chi-square; the left panel indicates the case of a fixed $H_0$, while in the right panel it has been analytically marginalized. The bottom panels show the one-dimensional posterior likelihoods for both parameters; the dashed lines correspond to the fixed $H_0$ and the solid lines to the marginalized case.}
\label{figure2}
\end{figure*}

\begin{figure*}
\centering
\includegraphics[width=7.5cm]{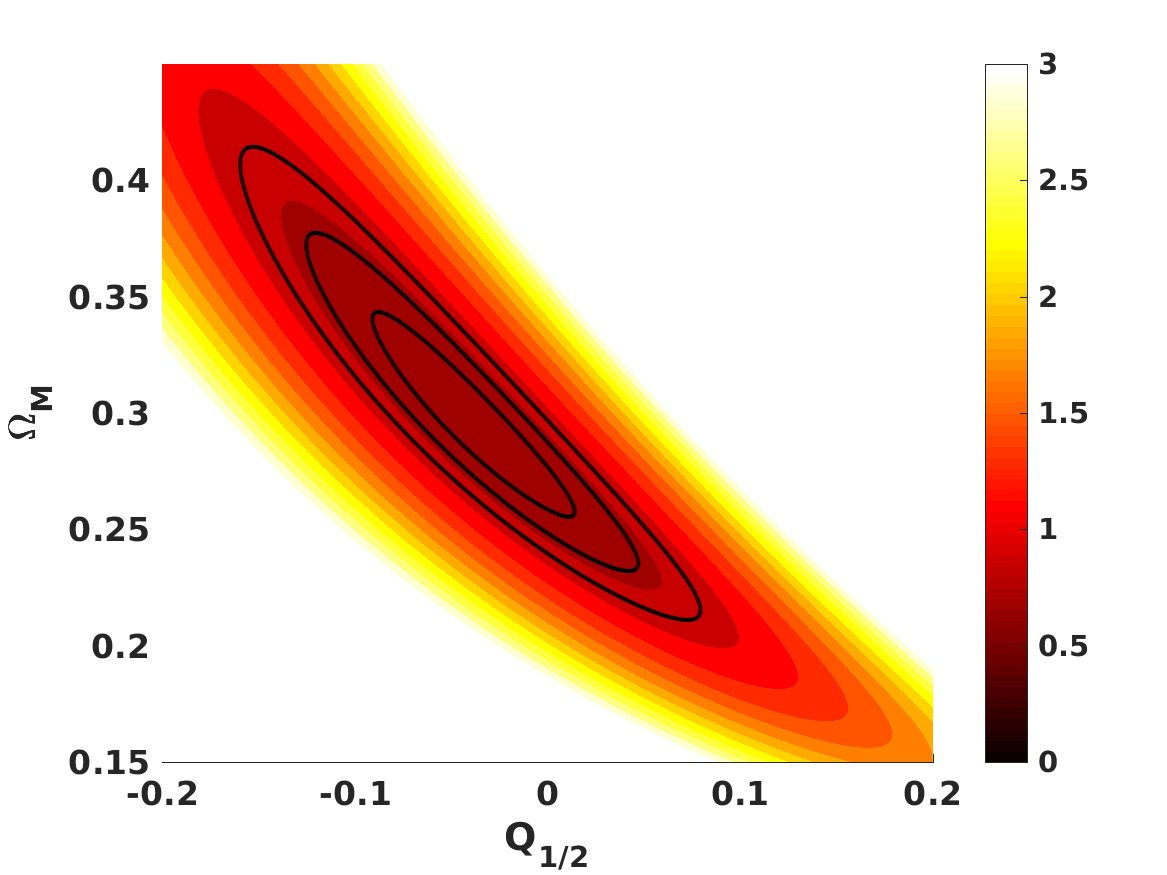}
\includegraphics[width=7.5cm]{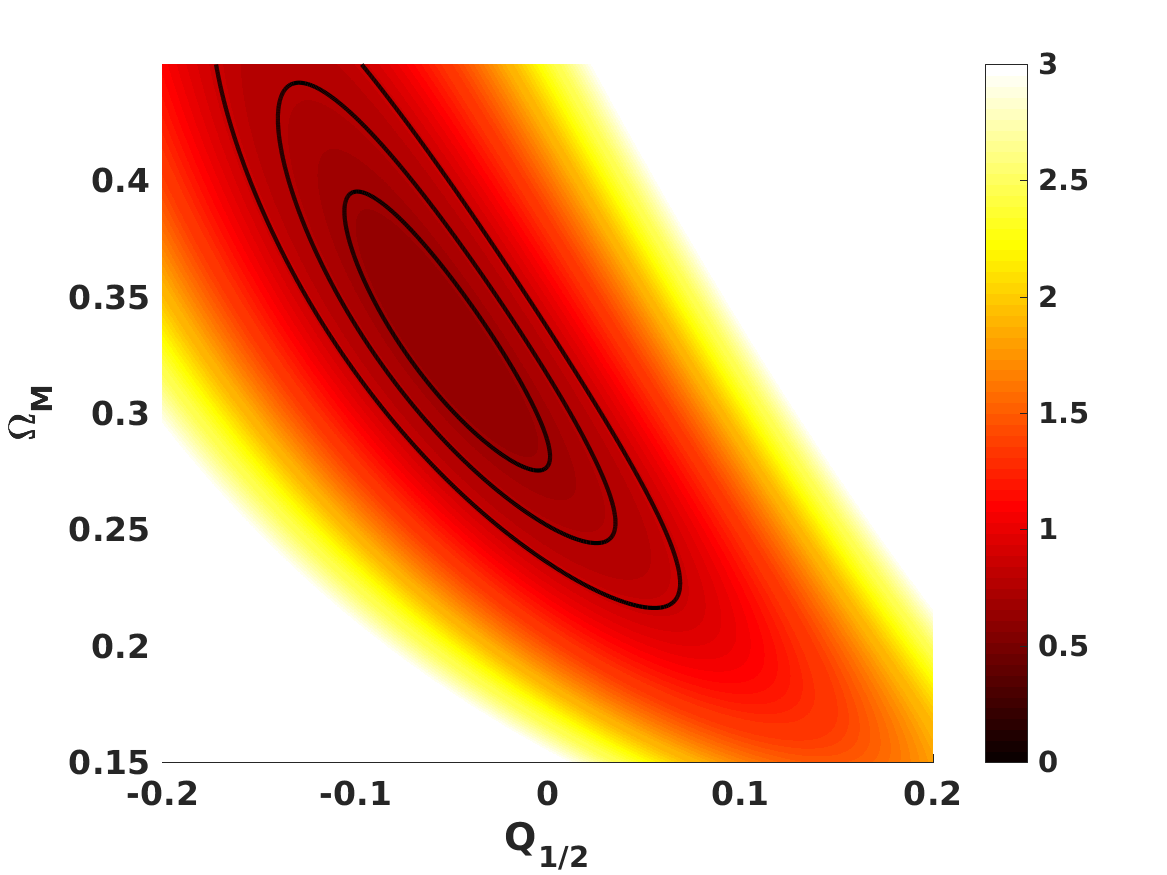}
\includegraphics[width=7.5cm]{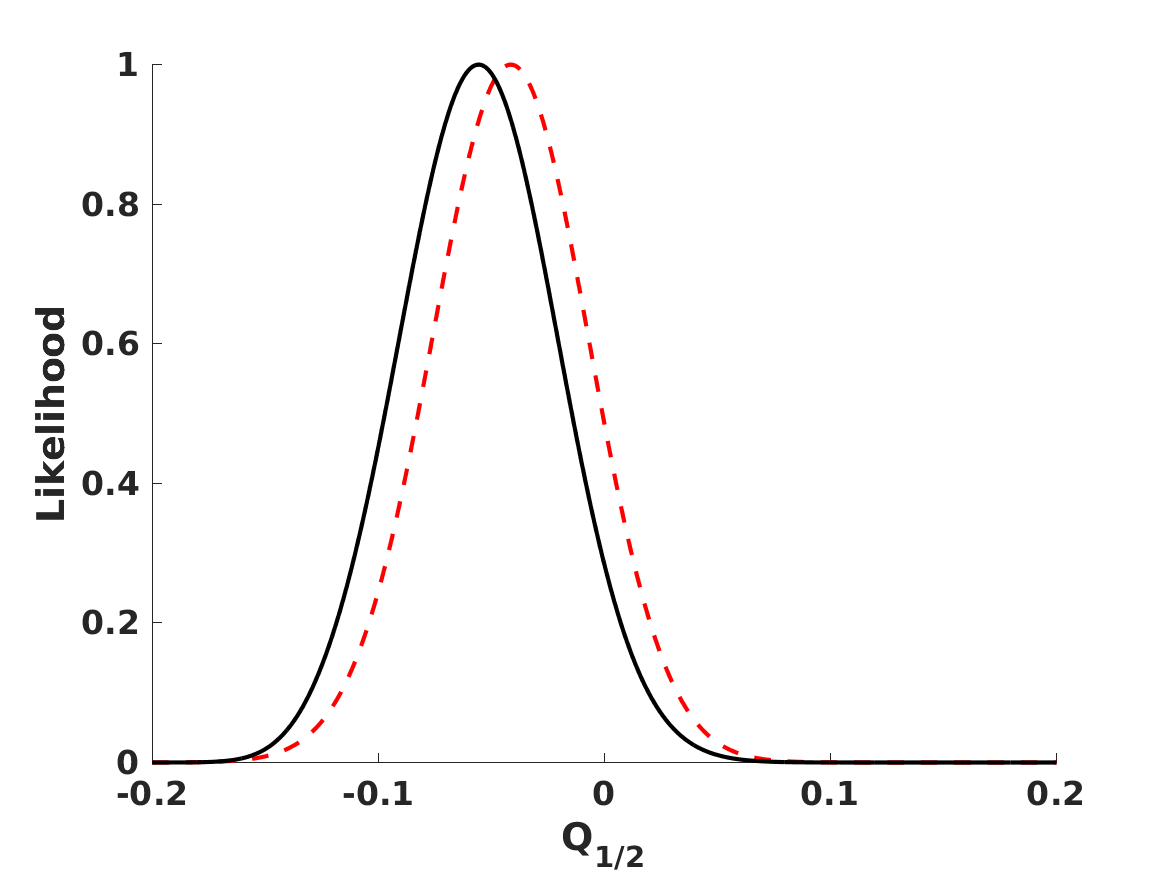}
\includegraphics[width=7.5cm]{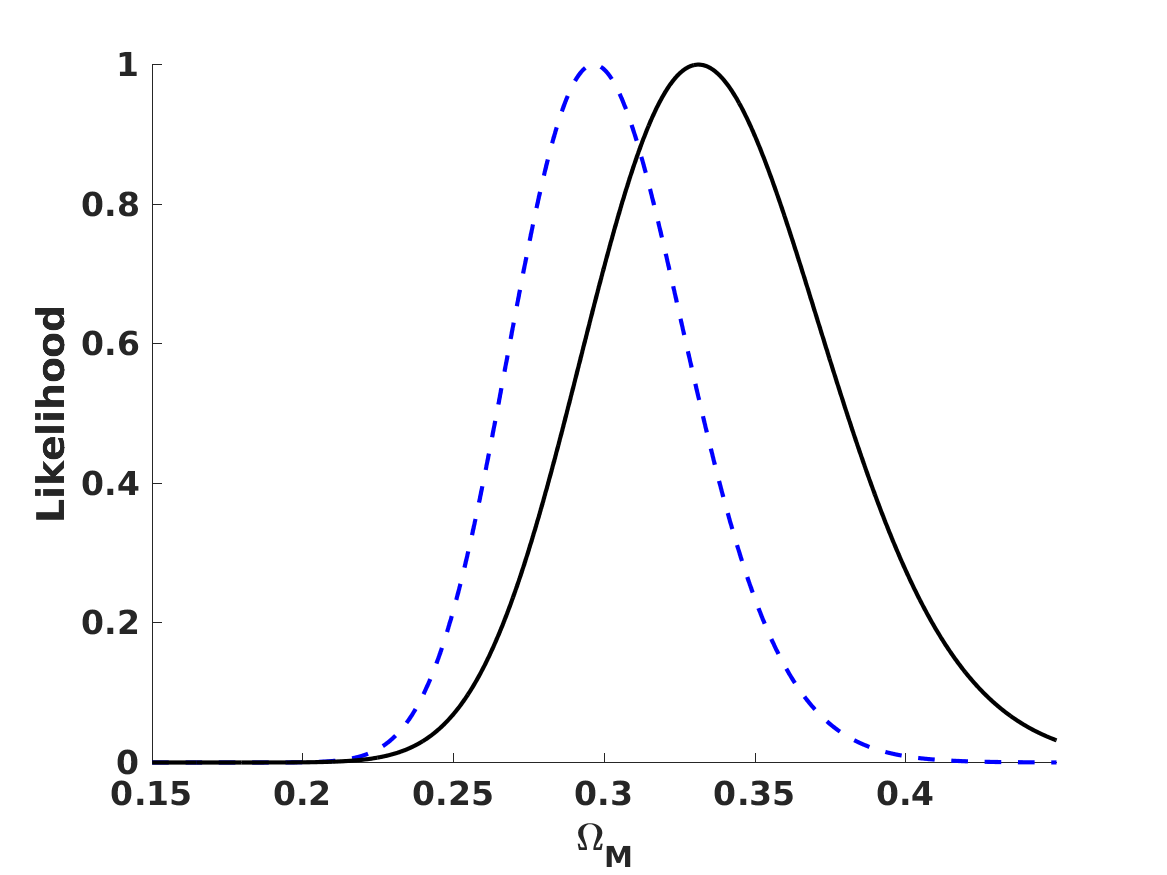}
\caption{Current constraints for the $n=1/2$ case, using $Q_{1/2}$ as free parameter. In the top panels the black solid curves show the one, two, and three sigma confidence levels in the two-dimensional plane, while the color map depicts the reduced chi-square; the left panel indicates the case of a fixed $H_0$, while in the right panel it has been analytically marginalized. The bottom panels show the one-dimensional posterior likelihoods for both parameters; the dashed lines correspond to the fixed $H_0$ and the solid lines to the marginalized case.}
\label{figure3}
\end{figure*}

We carried out a standard likelihood analysis with $\Omega_M$ and an additional $n$-dependent free parameter. As for the Hubble constant, $H_0$, we consider two different hypotheses: either this constant is fixed at $H_0=70$ km/s/Mpc, or it is analytically marginalized as discussed in \citet{Homarg}. The results of this analysis are summarized in Figs. \ref{figure1} and \ref{figure2} for the case $n=1$, and in Fig. \ref{figure3} for $n=1/2$, and in Table \ref{table1}, which lists the posterior likelihood constraints on the relevant parameters.

%%%%%%%%%%%%%%%%%%%%
%%%%%%%%%%%%%%%%%%%%

%%%%%%%%%%%%%%%%%%%%
%%%%%%%%%%%%%%%%%%%%

Starting with the $n=1$ case, the additional parameter is constrained to about the $10^{-3}$ level. Naturally the constraints on $\Omega_Q$ are tighter than those on $Q_1$ by about a factor of three, which is explained by the fact that $\Omega_M\sim0.3$; although we note that the degeneracy directions of the two-dimensional likelihoods are also slightly different. The choice of a fixed or marginalized Hubble constant has no significant impact on the constraints on the posterior likelihood for $\Omega_Q$ or $Q_1$, but does shift that of $\Omega_M$ by more than one standard deviation; a higher matter density is preferred in the marginalized case.

%%%%%%%%%%%%%%%%%%%%
%%%%%%%%%%%%%%%%%%%%

In the case $n=1/2$ the parameter $Q_{1/2}$ is constrained at the few percent level, and the anticorrelation between this parameter and the matter density is slightly higher than in the $n=1$ case. The preferred value for the matter density is also slightly higher than in the previous case, and again the main effect of the marginalization is to slightly increase the value of the preferred matter density. Although the model studied by \citet{Tutusaus} is somewhat different, we note that their parameter $\epsilon$ is analogous to our $Q_{1/2}$, and that their analysis (specifically for the Euclid galaxy clustering data with a fixed physical baryon density) leads to forecasts on low redshift constraints that are comparable to ours.

\begin{figure}
\centering
\includegraphics[width=8cm]{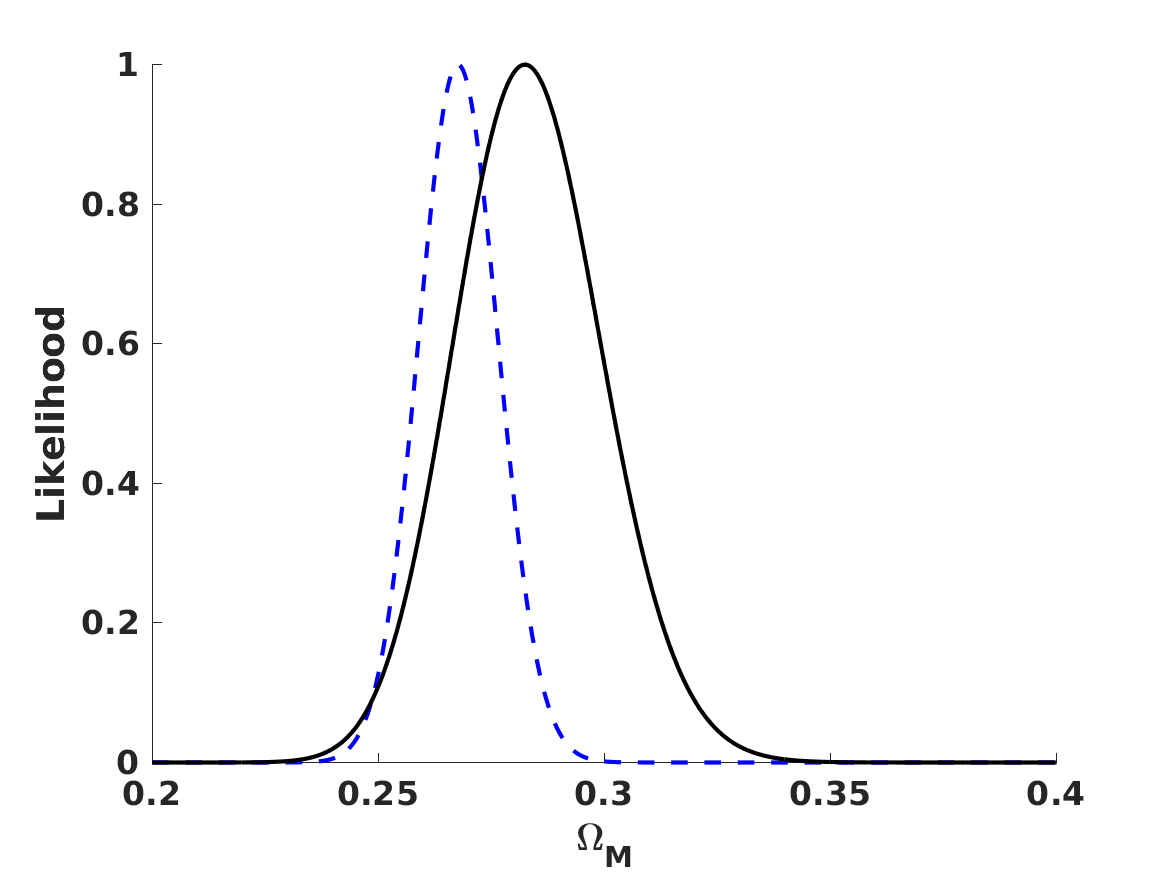}
\caption{One-dimensional posterior likelihood for the matter density from our supernova and Hubble parameter datasets for the $n=0$ case. The dashed line corresponds to the fixed $H_0$ and the solid line to the marginalized case.}
\label{figure4}
\end{figure}

For completeness we also show in Fig. \ref{figure4} and Table \ref{table1} the constraints on the matter density for the $n=0$ case, which corresponds to the flat $\Lambda$CDM case as we previously
mentioned. We see that the extension of the parameter space weakens the constraints on $\Omega_M$ in the $n=1/2$ case but not in the $n=1$ case. However, in all cases this extension shifts the preferred matter density to higher values.

%%%%%%%%%%%%%%%%%%%%
%%%%%%%%%%%%%%%%%%%%

%%%%%%%%%%%%%%%%%%%%%%%%%%%%%%%%%%%%%%%%%%%%%%%%%%%%%%%%%%%%%%%%%%%%%%%%%%%%%%
\begin{table}
\caption{One sigma posterior likelihoods on the matter density $\Omega_M$ and the additional free parameter for various flat models containing matter plus a cosmological constant, and a nonlinear matter Lagrangian with various different values of $n$. The constraints come from the combination of Pantheon supernova and Hubble parameter measurements, and are listed for fixed or analytically marginalized values of $H_0$.}
\label{table1}
\centering
\begin{tabular}{| c | c | c | c |}
\hline
Model & $H_0$ & Model parameter & $\Omega_M$ \\
\hline
$n=1$, $\Omega_Q$ & Fixed  & $(-0.7\pm0.5)\times10^{-3}$ & $0.28\pm0.01$  \\
$n=1$, $\Omega_Q$ & Marginalized  & $(-1.0\pm0.6)\times10^{-3}$ & $0.31\pm0.02$  \\
\hline
$n=1$, $Q_1$ & Fixed  & $(-2.6\pm1.7)\times10^{-3}$ & $0.28\pm0.01$  \\
$n=1$, $Q_1$ & Marginalized  & $(-3.3\pm1.8)\times10^{-3}$ & $0.31\pm0.02$  \\
\hline
$n=1/2$, $Q_{1/2}$ & Fixed  & $(-4.1\pm3.4)\times10^{-2}$ & $0.30\pm0.03$  \\
$n=1/2$, $Q_{1/2}$ & Marginalized  & $(-5.6\pm3.5)\times10^{-2}$ & $0.33\pm0.04$  \\
\hline
$n=0$ ($\Lambda$CDM) & Fixed  & N/A & $0.27\pm0.01$  \\
$n=0$ ($\Lambda$CDM) & Marginalized  & N/A & $0.28\pm0.02$  \\
\hline
\end{tabular}
\end{table}
%%%%%%%%%%%%%%%%%%%%%%%%%%%%%%%%%%%%%%%%%%%%%%%%%%%%%%%%%%%%%%%%%%%%%%%%%%%%%%

%%%%%%%%%%%%%%%%%%%%%%%%%%%%%%%%%%%%%%%%%%%%%%%%%%%%%%%%%%%%%%%%%%%%%%%%%%
\subsection{Robustness tests with other datasets}
\label{robust}

As a test of the robustness of the constraints obtained in the previous section, we repeat the analysis for some alternative choices of cosmological datasets. We focus on the three models described by parameters $Q_n$ (ignoring the case with $\Omega_Q$)

\begin{figure*}
\centering
\includegraphics[width=7.5cm]{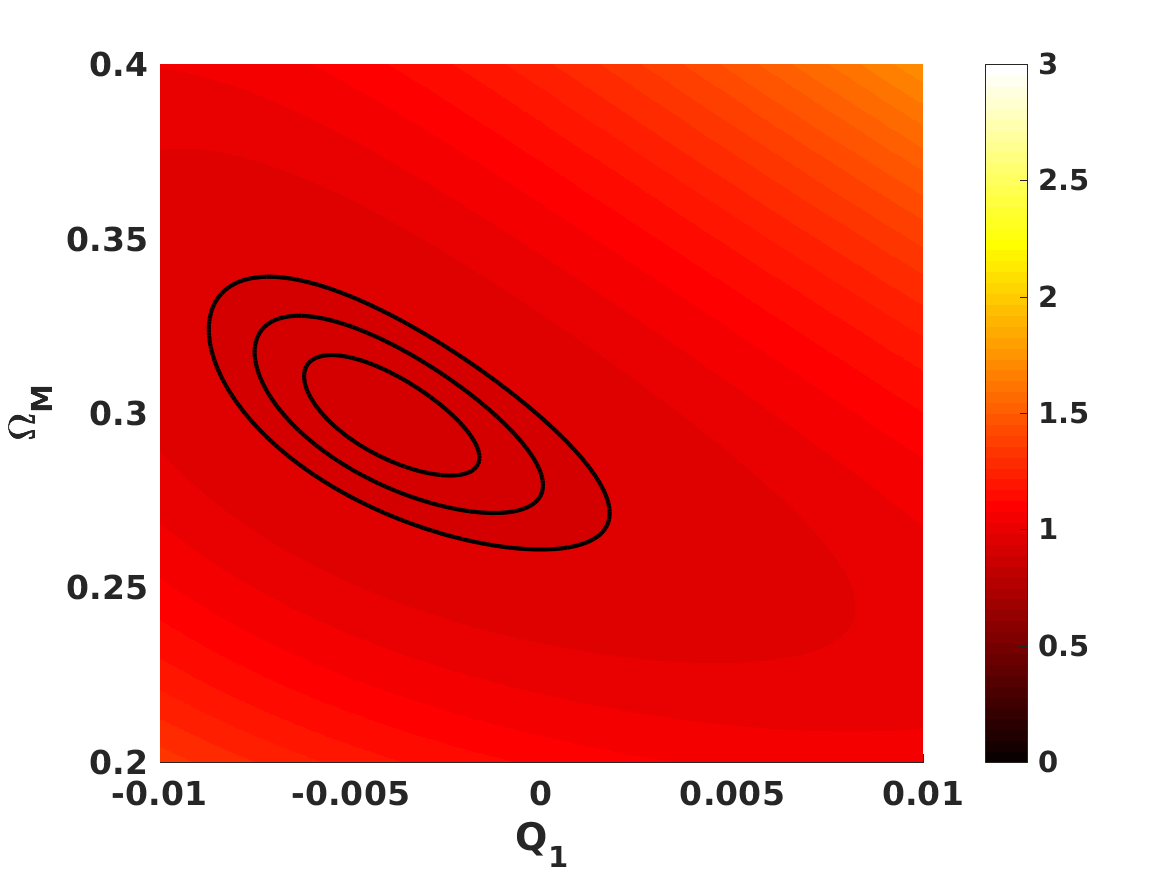}
\includegraphics[width=7.5cm]{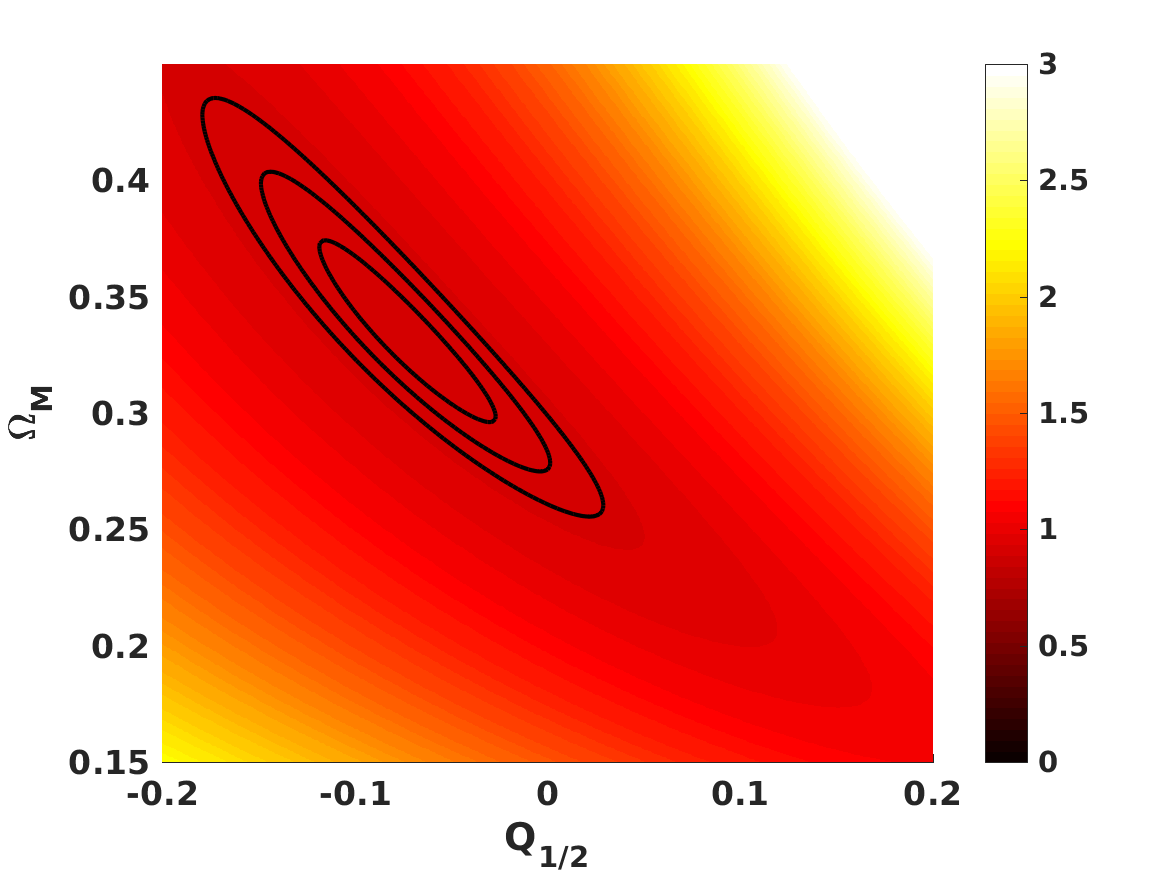}
\includegraphics[width=7.5cm]{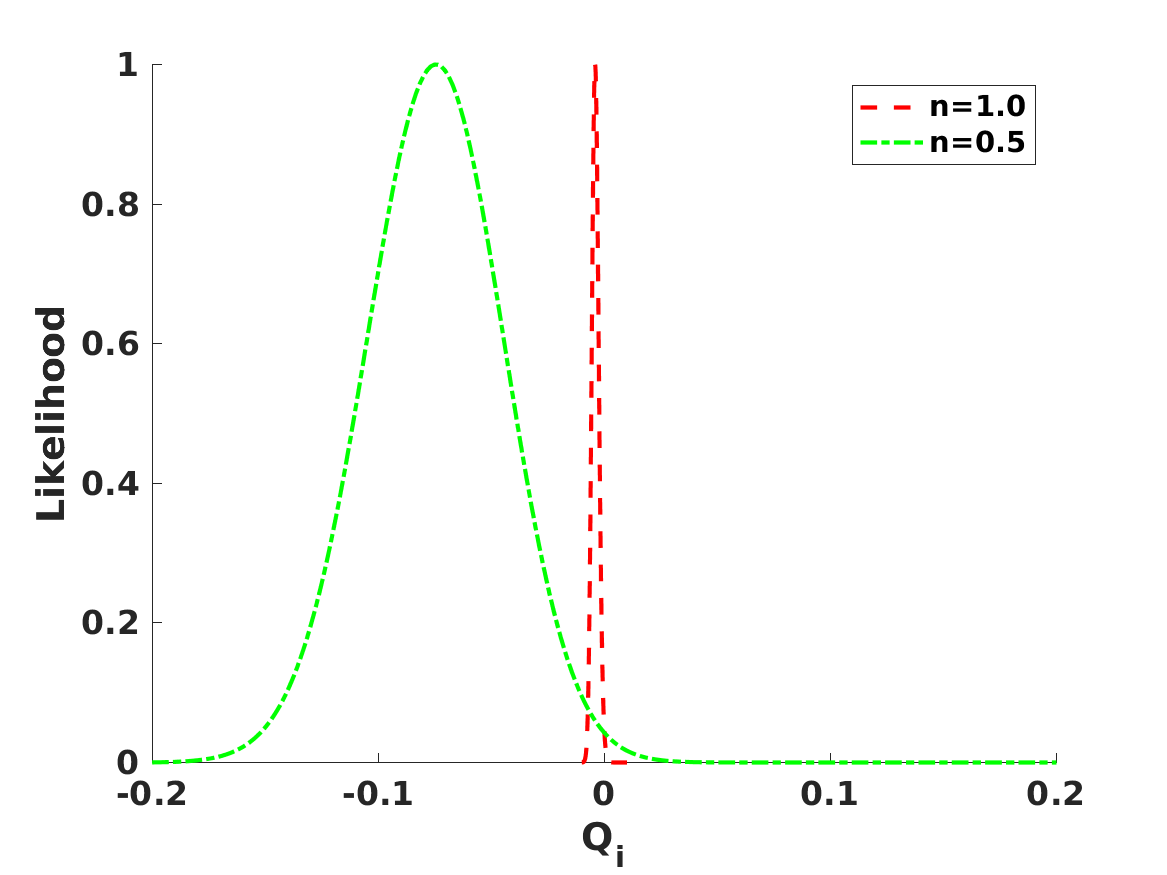}
\includegraphics[width=7.5cm]{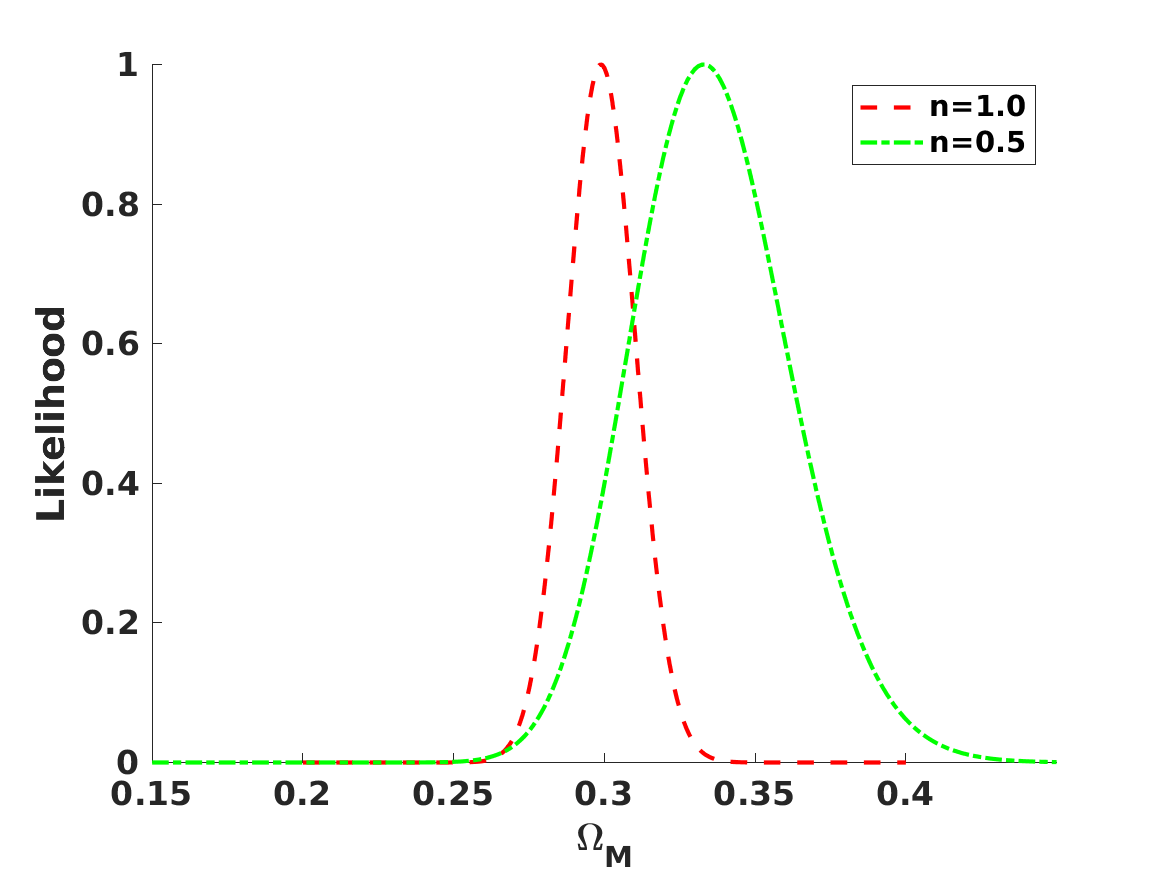}
\caption{Current constraints for various values of $n$, using the Union2.1 supernova data instead of the Pantheon data. In the top left ($n=1$) and top right ($n=1/2$) panels the black solid curves show the one, two, and three sigma confidence levels in the two-dimensional plane, while the color map depicts the reduced chi-square. The bottom panels show the one-dimension posterior likelihoods for the parameters $Q_1$ and $Q_{1/2}$ and the matter density $\Omega_M$ in both cases. The Hubble constant has been kept fixed throughout.}
\label{figure5}
\end{figure*}

Firstly, we use the Union2.1 Type Ia supernova compilation of \citet{Suzuki} instead of the Pantheon supernovas. This contains a total of 580 supernovas in the redshift range $0.015<z<1.414$. For simplicity we only carry out the analysis for a Hubble constant fixed at $H_0=70$ km/s/Mpc. The results are summarized in Fig. \ref{figure5} and in Table \ref{table2}.

%%%%%%%%%%%%%%%%%%%%
%%%%%%%%%%%%%%%%%%%%

We find that with the Union2.1 data the constraints on the parameter $Q_n$ are slightly stronger for both $n=1$ and $n=1/2$, while those on $\Omega_M$ are almost identical. There is good overall consistency in all cases, the one point that is perhaps worthy of notice is that with the Union2.1 data the best-fit values for $Q_n$ are shifted to more negative values by about one standard deviation, while those of the matter density are correspondingly shifted to higher values.

%%%%%%%%%%%%%%%%%%%%%%%%%%%%%%%%%%%%%%%%%%%%%%%%%%%%%%%%%%%%%%%%%%%%%%%%%%%%%%
\begin{table}
\caption{One sigma posterior likelihoods on the matter density $\Omega_M$ and the additional free parameter $Q_n$ for various flat models containing matter plus a cosmological constant, and a nonlinear matter Lagrangian with various different values of $n$. The constraints come from the combination of Union2.1 supernova and Hubble parameter measurements with a fixed  $H_0=70$ km/s/Mpc.}
\label{table2}
\centering
\begin{tabular}{| c | c | c |}
\hline
Model & $Q_n$ & $\Omega_M$ \\
\hline
$n=1$ & $(-4.1\pm1.5)\times10^{-3}$ & $0.30\pm0.01$  \\
\hline
$n=1/2$ & $(-7.5\pm3.0)\times10^{-2}$ & $0.33\pm0.03$  \\
\hline
\end{tabular}
\end{table}
%%%%%%%%%%%%%%%%%%%%%%%%%%%%%%%%%%%%%%%%%%%%%%%%%%%%%%%%%%%%%%%%%%%%%%%%%%%%%%

We also test replacing the Hubble parameter measurements with a prior on the matter density. One motivation for doing this is that this compilation of Hubble parameter measurements is heterogeneous. Some of the measurements come from galaxy clustering and baryon acoustic oscillations observations, while others come from the so-called cosmic chronometers or differential age method proposed by \citet{Jimenez}, and it is not currently clear that possible systematics issues of the cosmic chronometers method are well understood and under control \citep{LiuLu,Vazdekis,Concas,Corredoira}. Our matter priors are provided by the Planck satellite \citep{Planck}, specifically $\Omega_M=0.315\pm0.007$, or from the Dark Energy Survey (DES) Year 1 galaxy clustering and weak gravitational lensing data, which is referred to by the DES collaboration as the $3\times2$ data \citep{DESY1}, specifically $\Omega_M=0.28\pm0.04$. 

%%%%%%%%%%%%%%%%%%%%
\begin{figure*}
\centering
\includegraphics[width=7.5cm]{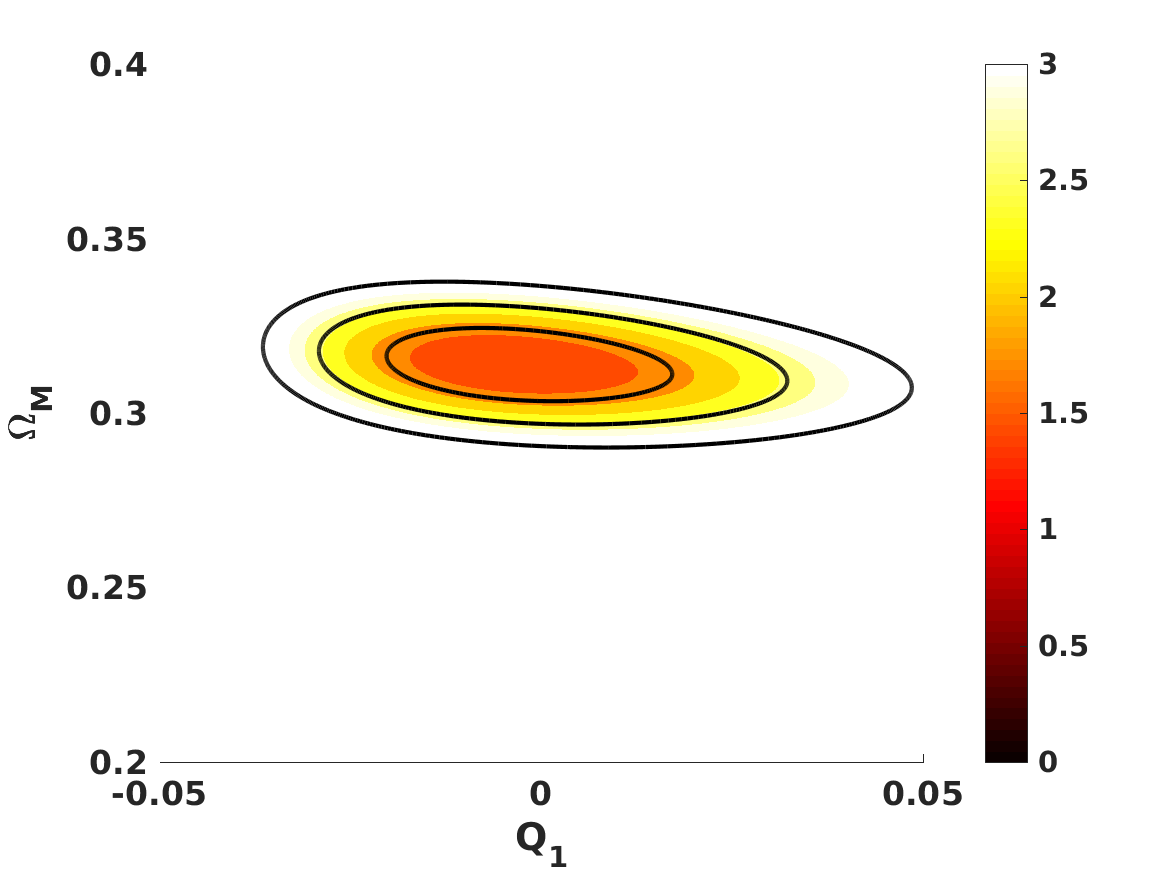}
\includegraphics[width=7.5cm]{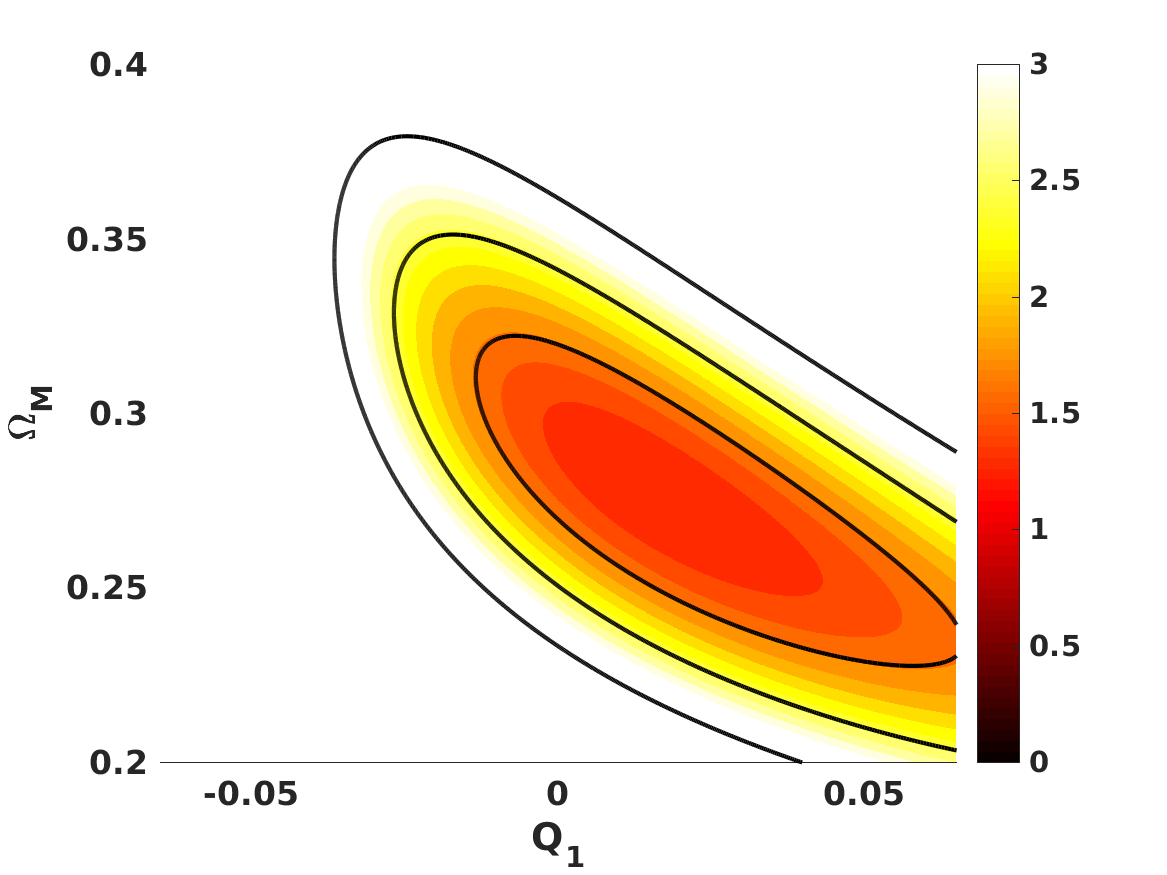}
\includegraphics[width=7.5cm]{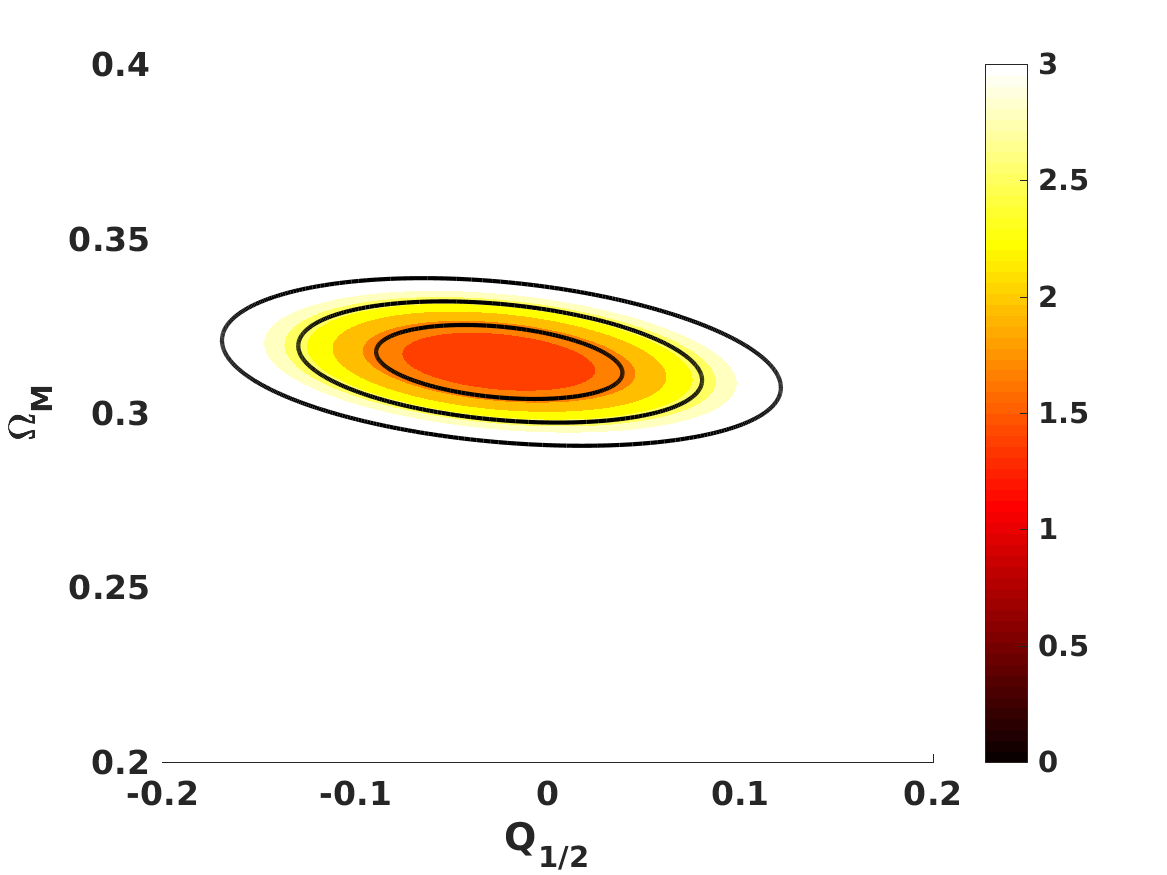}
\includegraphics[width=7.5cm]{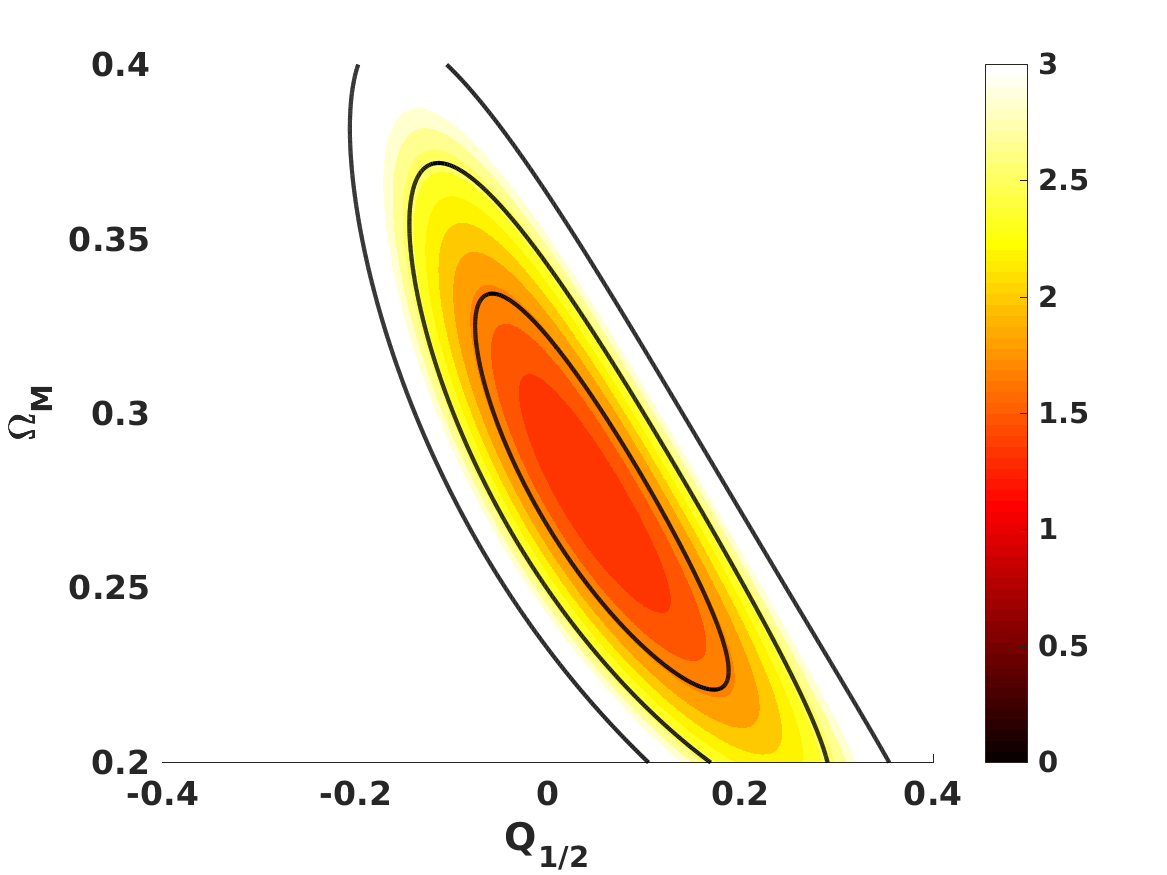}
\caption{Current constraints for various values of $n$, using a prior on the matter density instead of the Hubble parameter data. The left panels show the results for a Planck prior, while the right panels show the results for a DES prior. The black solid curves show the one, two, and three sigma confidence levels in the two-dimensional plane, while the color map depicts the reduced chi-square. The top and bottom panels correspond to $n=1$ and $n=1/2$, respectively.}
\label{figure6}
\end{figure*}
%%%%%%%%%%%%%%%%%%%%

The results of this analysis are summarized in Fig. \ref{figure6} and in Table \ref{table3}. Compared to our baseline constraints, the Planck prior leads to constraints for $Q_1$, which are weaker by about a factor of seven and by about $20\%$ for $Q_{1/2}$. On the other hand, the DES prior leads to constraints that are about a factor of two weaker than those of Planck for each of the models.

%%%%%%%%%%%%%%%%%%%%%%%%%%%%%%%%%%%%%%%%%%%%%%%%%%%%%%%%%%%%%%%%%%%%%%%%%%%%%%
\begin{table}
\caption{One sigma posterior likelihoods on the additional free parameter $Q_n$ for various flat models containing matter plus a cosmological constant, and a nonlinear matter Lagrangian with various different values of $n$. The constraints come from the combination of the Pantheon supernova data with the Planck or DES priors described in the main text.}
\label{table3}
\centering
\begin{tabular}{| c | c | c |}
\hline
Model parameter & Planck & DES \\
\hline
$Q_1$ & $(-0.4\pm1.2)\times10^{-2}$ & $(1.6^{+2.9}_{-2.1})\times10^{-2}$ \\
\hline
$Q_{1/2}$ & $(-2.6\pm4.2)\times10^{-2}$ & $(3.8^{+9.0}_{-8.0})\times10^{-2}$ \\
\hline
\end{tabular}
\end{table}
%%%%%%%%%%%%%%%%%%%%%%%%%%%%%%%%%%%%%%%%%%%%%%%%%%%%%%%%%%%%%%%%%%%%%%%%%%%%%%

%%%%%%%%%%%%%%%%%%%%%%%%%%%%%%%%%%%%%%%%%%%%%%%%%%%%%%%%%%%%%%%%%%%%%%%%%%
\section{Forecasts for specific values of $n$}
\label{frcst}

Before moving on to the broader parameter space of these models, we briefly discuss how the constraints discussed in the previous sections might be improved by future observations. Specifically, we consider measurements of the redshift drift by the ELT \citep{Liske,HIRES}, which will directly probe the expansion of the universe in the deep matter era, and an improved supernova dataset.

The redshift drift of an astrophysical object following the cosmological expansion can be shown to be given by \citep{Sandage}
\be
\Delta z=\tau_{obs} H_0 \left[1+z-E(z)\right]\,,
\ee
where $\tau_{obs}$ is the observation time span, although the actual observable is a spectroscopically measured velocity
\be
\Delta v=k\tau_{obs}h\left[1-\frac{E(z)}{1+z}\right]\,,
\ee
which for convenience we expressed in terms of $E(z)$ and $h$; we also introduced the normalization constant $k$, which has the value $k=3.064$ cm/s if $\tau_{obs}$ is expressed in years. In the case of the ELT, and specifically its planned high-resolution spectrograph currently known as ELT-HIRES \citep{HIRES}, the uncertainty in the velocity measurement is expected to be \citep{Liske}
\be
\sigma_v(z)=1.35\left(\frac{2370}{S/N}\right)\sqrt{\frac{30}{N_{qso}}}\left(\frac{5}{1+z_{qso}}\right)^\lambda\,,
\ee
where $S/N$ denotes the signal to noise of the spectra available at the redshift bin $z_{qso}$ and $N_{qso}$ is the number of quasars observed at that redshift. The exponent of the last term is $\lambda=1.7$ for $z_{qso}\le4$ and $\lambda=0.9$ for $z_{qso}>4$.

\begin{figure*}
\centering
\includegraphics[width=7.5cm]{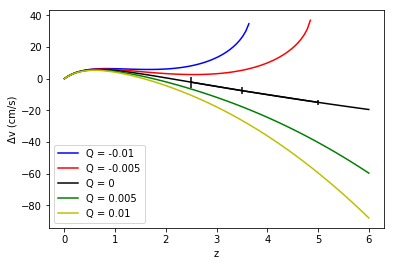}
\includegraphics[width=7.5cm]{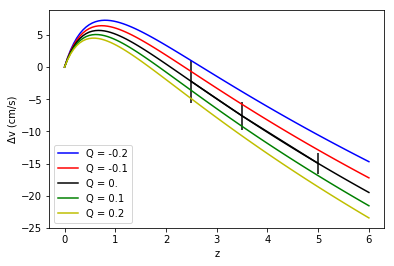}
\caption{Redshift dependence of the redshift drift signal (expressed as the corresponding spectroscopic velocity) for $n=1$ in the left panel and $n=1/2$ in the right panel, and various model parameters $Q_n$. The error bars expected from future ELT measurements are also depicted.}
\label{figure7}
\end{figure*}

We assume a realistic observation program with a time span of $\tau_{obs}=20$ years, a signal to noise $S/N=3000$ in each measurement, and three different measurements at redshift bins centered at $z_{qso}=2.5, 3.5, 5.0$, each based on data from $N_{qso}=10$ quasars. The predicted signals for the two representative models ($n=1$ and $n=1/2$) are shown in Fig. \ref{figure7}, for various choices of the free model parameter $Q_n$ and fixed values $\Omega_M=0.3$ and $h=0.7$. It is clear that the constraining power of the redshift drift for constraining these models---as well as for many others---lies at the high end of the redshift range, both because the measurements are potentially more sensitive and because deviations from the standard $\Lambda$CDM behavior are larger there. This is especially visible for the $n=1$ case, since values of $Q_1<0$ lead to a very unorthodox high redshift behavior.

%%%%%%%%%%%%%%%%%%%%
%%%%%%%%%%%%%%%%%%%%

The work of \citet{Riess} also discusses a future dataset of supernova measurements from the proposed WFIRST satellite \citep{WFIRST}. Their analysis leads to the following values for percent errors on $E(z)$:  $\sigma=1.3, 1.1, 1.5, 1.5, 2.0, 2.3, 2.6, 3.4, 8.9$ for the nine redshift bins centered at $z=0.07, 0.20, 0.35, 0.60, 0.80, 1.00, 1.30, 1.70, 2.50$, respectively. These measurements are not fully independent, as there are pairwise correlations among some of the measurements, but these effects are small. Their simulated dataset was also obtained under the assumption of a flat universe.

We thus forecast constraints on the two fixed-$n$ models introduced in Sect. \ref{mods} and constrained with current data in Sects. \ref{cnstr}-\ref{robust} for a combined mock dataset of ELT redshift drift and WFIRST supernova measurements. In addition to this dataset on its own, we will also consider a further case where the dataset is complemented by a Planck-like prior that has an uncertainty on the matter density $\sigma(\Omega_m)=0.007$. In all cases our fiducial will be a flat $\Lambda$CDM model with $\Omega_m=0.3$ and $h=0.7$.

\begin{figure*}
\centering
\includegraphics[width=7.5cm]{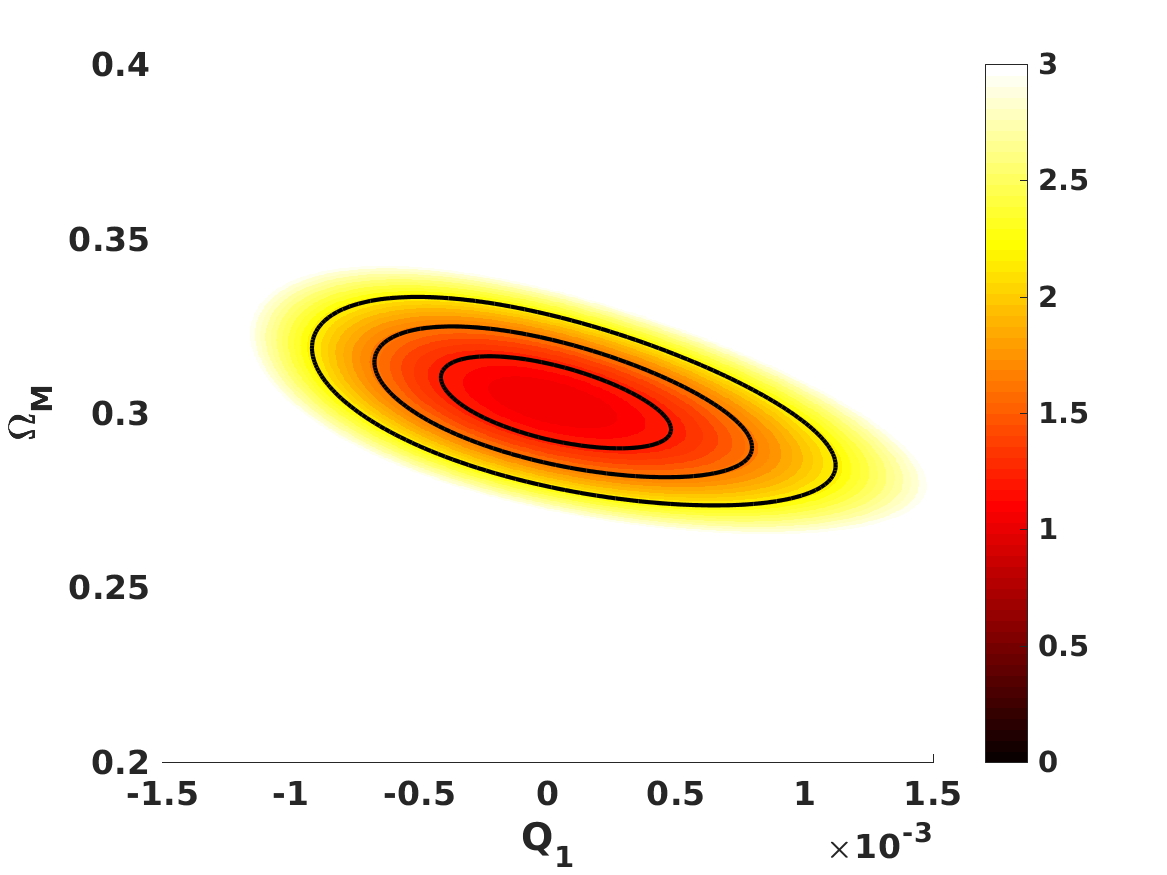}
\includegraphics[width=7.5cm]{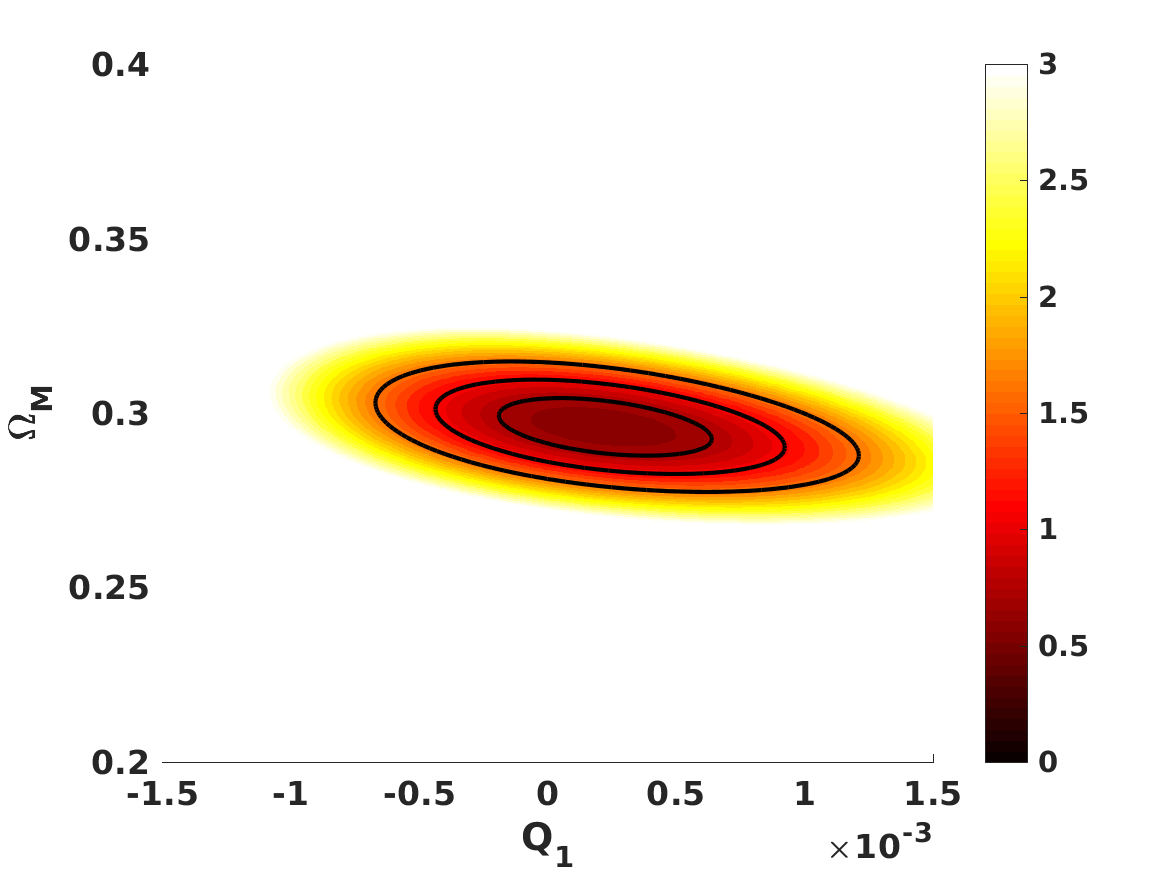}
\includegraphics[width=7.5cm]{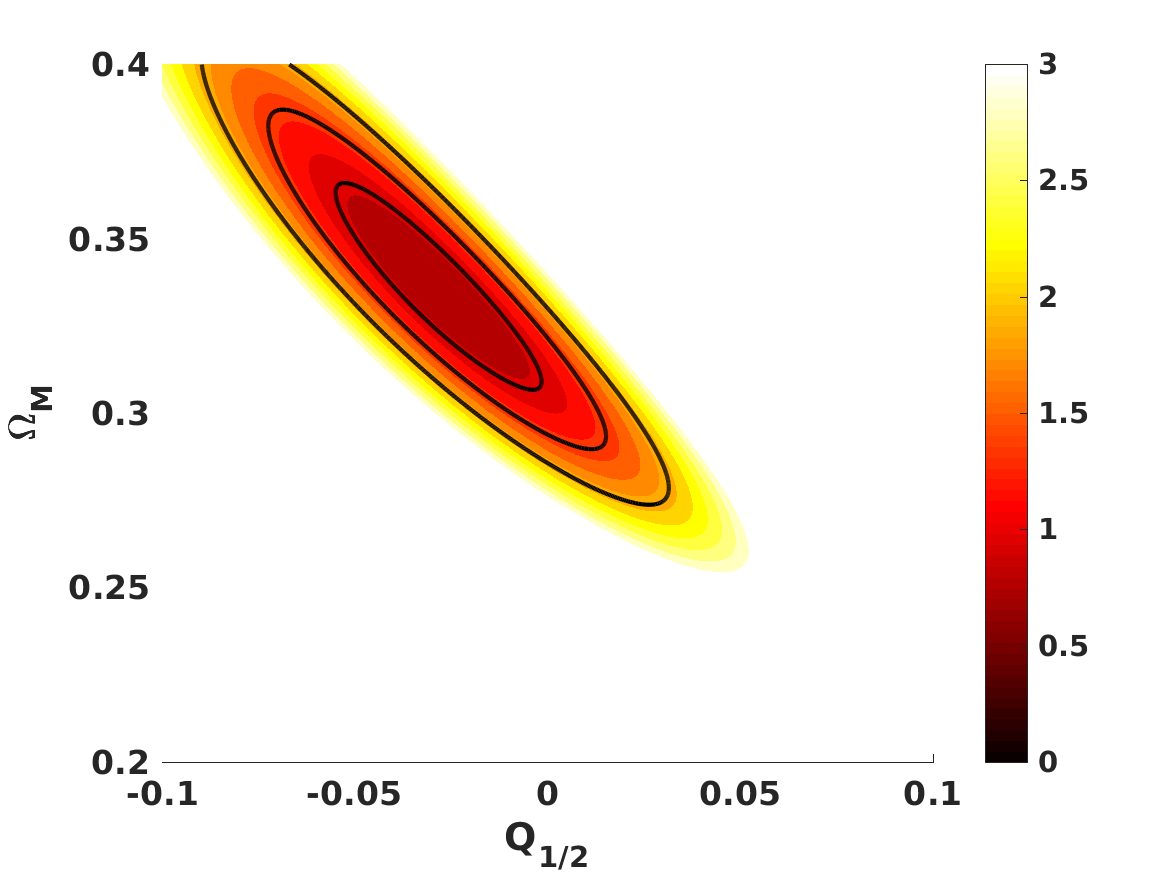}
\includegraphics[width=7.5cm]{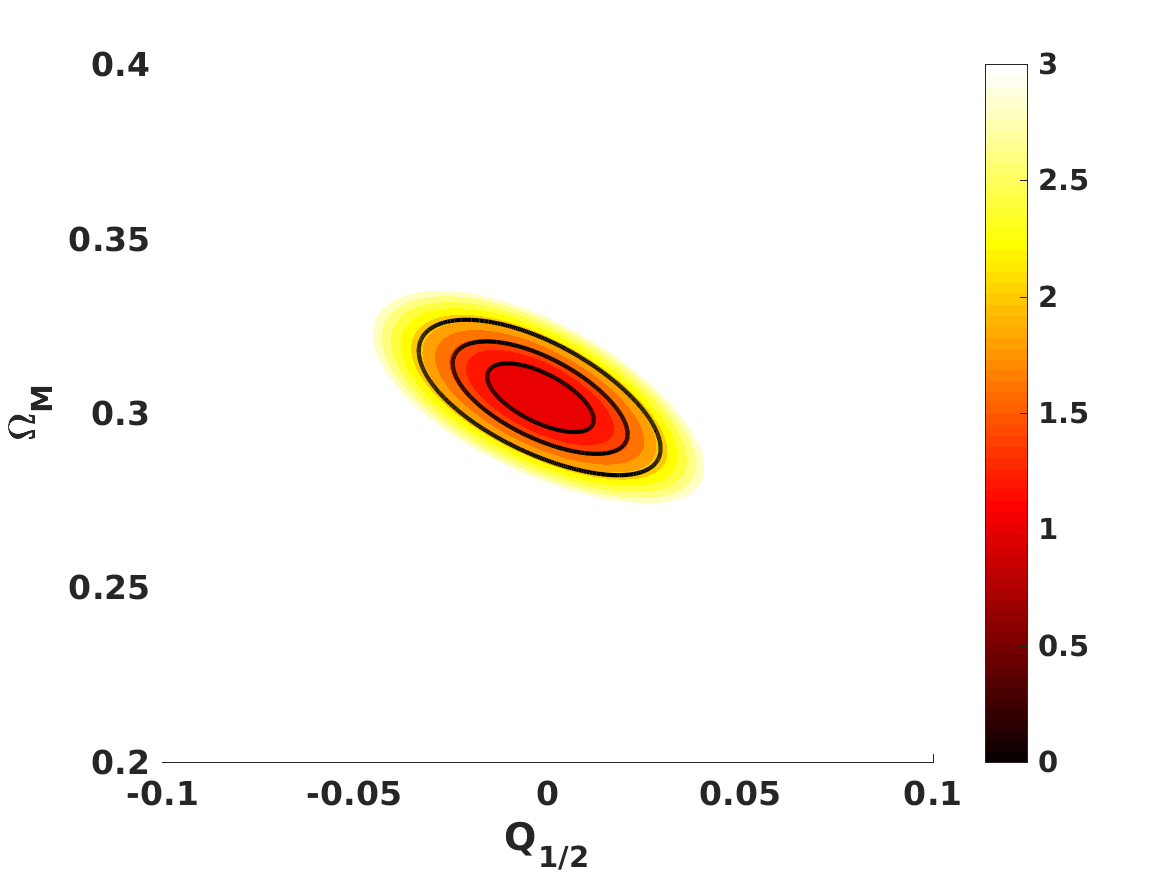}
\caption{Forecasted constraints in the $Q_n$--$\Omega_M$ plane for two choices of $n$. The black solid curves show the one, two, and three sigma confidence levels in the two-dimensional plane, while the color map depicts the reduced chi-square. The top and bottom panels correspond to $n=1$ and $n=1/2$, respectively. The left side panels show the forecasts for the combination of simulated ELT redshift drift and WFIRST supernova measurements, while for the right side panels a Planck-like prior on the matter density has been added to this data.}
\label{figure8}
\end{figure*}

The results of this analysis are summarized in Fig. \ref{figure8} and in Table \ref{table4}, which compares the expected uncertainty on the parameters $Q_n$ with those obtained under various assumptions and discussed in previous sections. We note the significant improvements in all cases, by at least a factor of two, and in some cases more than one order of magnitude. As previously mentioned, it is also clear that the redshift drift strongly excludes values of $Q_1<0$.

%%%%%%%%%%%%%%%%%%%%
%%%%%%%%%%%%%%%%%%%%

%%%%%%%%%%%%%%%%%%%%%%%%%%%%%%%%%%%%%%%%%%%%%%%%%%%%%%%%%%%%%%%%%%%%%%%%%%%%%%
\begin{table}
\caption{One-sigma uncertainty (marginalizing over other relevant parameters) for the additional free parameter, $Q_n$ of two flat fixed-$n$ models, including a cosmological constant, for various combinations of observational data. The first three rows refer to current data, while the last two are forecasts for future data.}
\label{table4}
\centering
\begin{tabular}{| c | c | c c |}
\hline
Data & Planck prior & $\sigma(Q_1)$ & $\sigma(Q_{1/2})$ \\
\hline
Pantheon$+$Hubble & No & $1.8\times10^{-3}$ & $3.5\times10^{-2}$ \\
Union2.1$+$Hubble & No & $1.5\times10^{-3}$ & $3.0\times10^{-2}$ \\
Pantheon & Yes & $1.2\times10^{-2}$ & $4.2\times10^{-2}$  \\
\hline
ELT$+$WFIRST & No & $3.0\times10^{-4}$ & $1.8\times10^{-2}$ \\
ELT$+$WFIRST & Yes & $2.7\times10^{-4}$ & $9.2\times10^{-3}$ \\
\hline
\end{tabular}
\end{table}
%%%%%%%%%%%%%%%%%%%%%%%%%%%%%%%%%%%%%%%%%%%%%%%%%%%%%%%%%%%%%%%%%%%%%%%%%%%%%%

%%%%%%%%%%%%%%%%%%%%%%%%%%%%%%%%%%%%%%%%%%%%%%%%%%%%%%%%%%%%%%%%%%%%%%%%%%
\section{Cosmological evolution and constraints for generic $n$}
\label{genn}

So far we studied particular cases of this class of models corresponding to specific (fixed) values of the power $n$ of the nonlinear matter Lagrangian. We now relax the $n=const.$ assumption and study the more general case in which $n$ is itself a model parameter that is allowed to vary and constrained by observations. In order to do this we define a dimensionless cosmological density $r$, via $\rho = r \rho_0$, where $\rho_0$ is the present day density, as well as a generic parameter
\be
Q=\frac{\eta}{\kappa}\rho_0^{2n-1}\,.
\ee
With these assumptions the continuity equation for a model containing matter and possibly also a cosmological constant, previously introduced in Eq. \ref{maineq2}, and expressed in terms of redshift and not time, has the form
\be
\frac{dr}{dz}=\frac{3r}{1+z}\times\frac{1+nQr^{2n-1}}{1+(2n-1)nQr^{2n-1}}\,;
\ee
this can be numerically integrated to yield $r(z)$, and the result can then be substituted into the Friedmann equation, previously defined in Eq. \ref{maineq1}, which in these variables becomes
\be
E^2(z)=\frac{H^2(z)}{H_0^2}=\Omega_\Lambda+\Omega_Mr+\left(n-\frac{1}{2}\right)Q\Omega_Mr^{2n}\,.
\ee
Our flatness assumption requires that $\Omega_\Lambda=1-\Omega_M[1+(n-1/2)Q]$, and therefore the Friedmann equation becomes
\be
E^2(z)=\frac{H^2(z)}{H_0^2}=\Omega_\Lambda+\Omega_Mr+(1-\Omega_M-\Omega_\Lambda)r^{2n}\,.
\ee
Alternatively, we can also use the flatness assumption to elliminate $Q$ in the continuity equation, writing it as
\be
\frac{dr}{dz}=\frac{3r}{1+z}\times\frac{(2n-1)\Omega_M+2n(1-\Omega_M)r^{2n-1}}{(2n-1)[\Omega_M+2n(1-\Omega_M)r^{2n-1}]}\,;
\ee
in the special case $n=1/2$ we have the analytic solution already described in Section \ref{mods2}.

This shows that in a phenomenological sense these models could explain the recent acceleration of the universe without invoking a cosmological constant but relying instead on the nonlinearities of the matter Lagrangian in a matter-only universe with $n=0$. However, we note that even if these models can lead to accelerating universes without a cosmological constant, this is not per se sufficient to solve the ``old'' cosmological constant problem of why it should be zero.

For $n$ close to but not equal two zero, two things happen: on the one hand the (formerly) constant term  in the Friedmann equation becomes lowly varying, and on the other hand the continuity equation implies that the matter density does not behave exactly as $r\propto (1+z)^3$. We now discuss the extent to which deviations from the $n=0$ case are observationally allowed, and also the most general parameter space where a standard cosmological constant is also allowed.

%%%%%%%%%%%%%%%%%%%%
\begin{figure*}
\centering
\includegraphics[width=7.5cm]{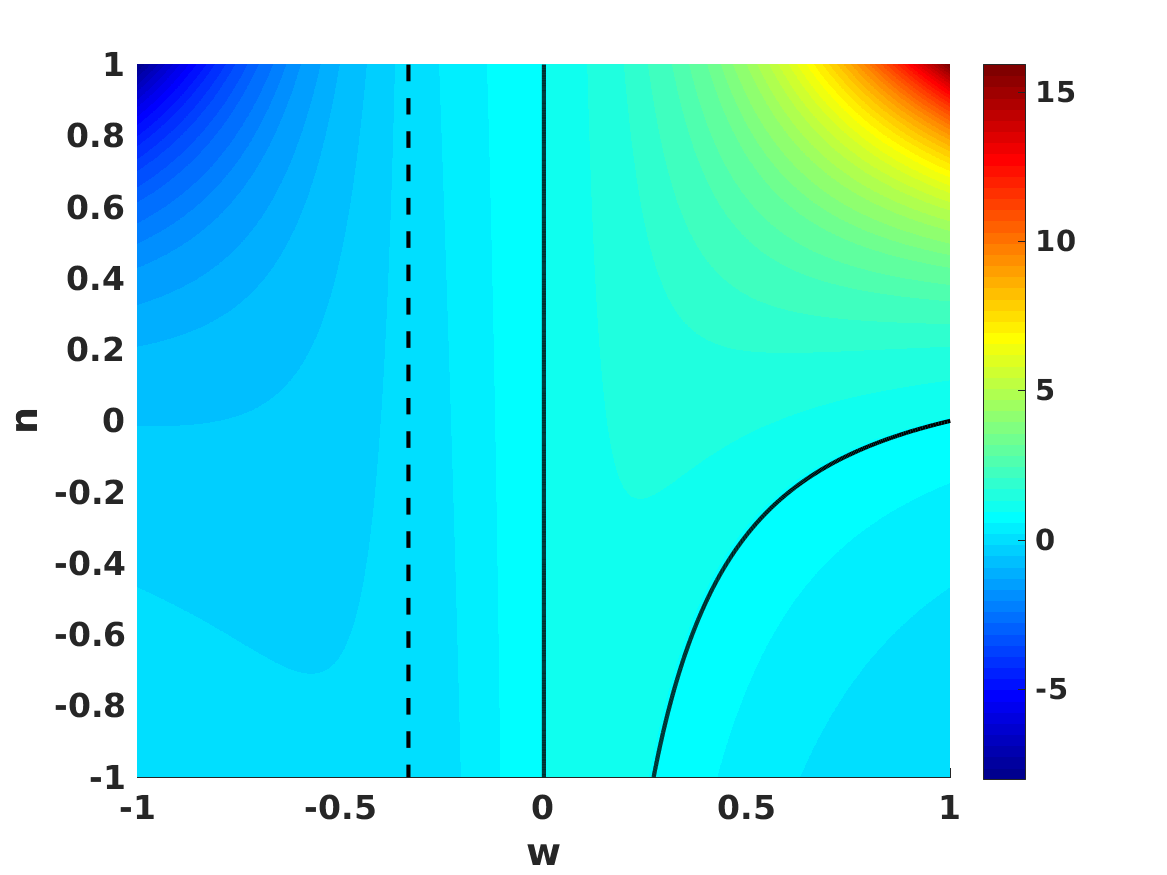}
\includegraphics[width=7.5cm]{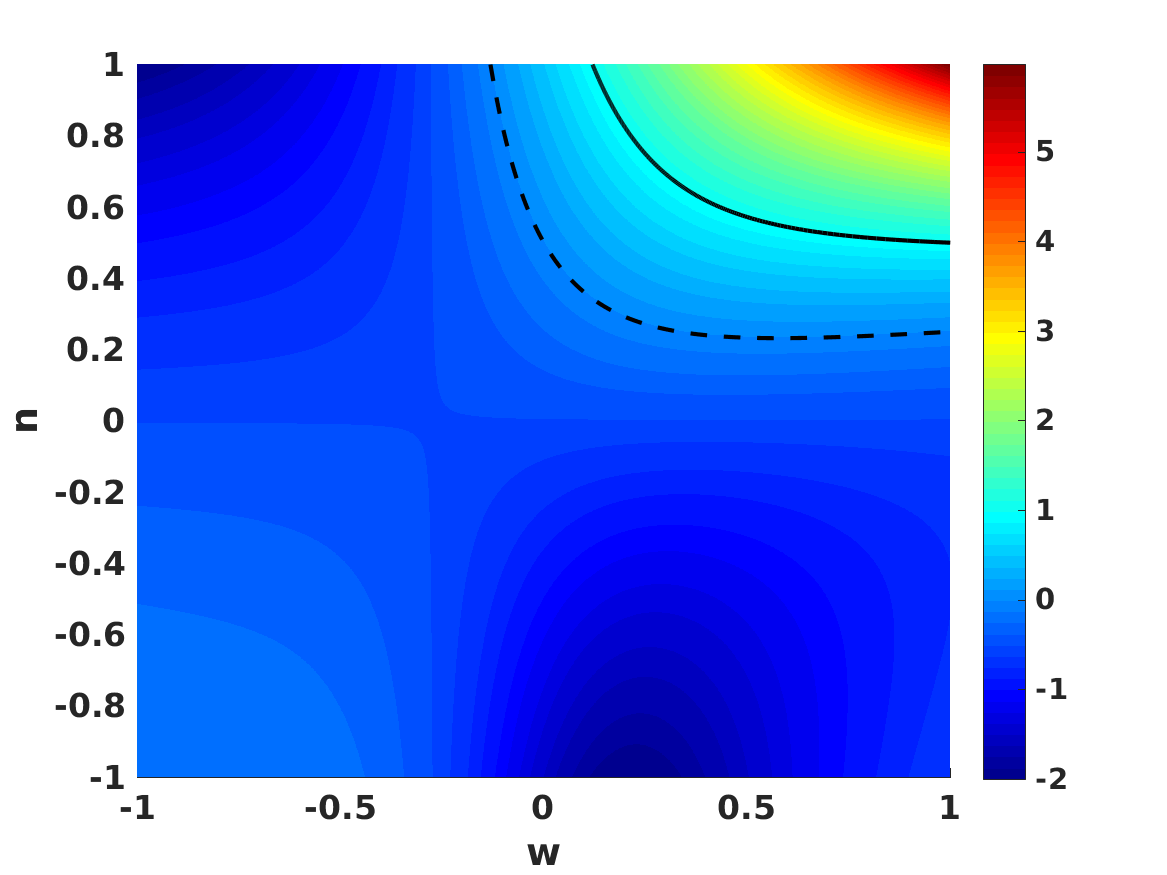}
\caption{Values of the functions $f_1$ and $f_2$ (cf. Eqs. \ref{deff1} and \ref{deff2}, respectively) as functions of the power $n$ of the nonlinear term and the dark energy equation of state $w$. The dashed and solid lines identify the locus of parameters where the functions have values of zero and unity, respectively. We note that the color bars in the two panels have different scales.}
\label{figure9}
\end{figure*}
%%%%%%%%%%%%%%%%%%%%

It is in fact possible to consider a further generalization: instead of considering a universe with a matter fluid, we can assume that this fluid has a constant equation of state $w=p/\rho=const.$ (with the matter case corresponding to $w=0$). In this case the continuity equation becomes
\be
\frac{dr}{dz}=\frac{3r}{1+z}(1+w)\times\frac{1+nQf_1(n,w)r^{2n-1}}{1+2nQf_2(n,w)r^{2n-1}}\,,
\ee
where for convenience we defined
\be
f_1(n,w)=(1+3w)(1+3w^2)^{n-1}\,, \label{deff1}
\ee
and
\be
f_2(n,w)=(1+3w^2)^{n-1}\left[\left(n-\frac{1}{2}\right)(1+3w^2)+4nw\right]\,. \label{deff2}
\ee
Figure \ref{figure9} illustrates the dependence of these functions on their two parameters. In this case the Friedmann equation has the form
\be
E^2(z)=\frac{H^2(z)}{H_0^2}=\Omega_\Lambda+\Omega_Mr+f_2(n,w)Q\Omega_Mr^{2n}\,,
\ee
with the flatness condition requiring $\Omega_\Lambda=1-\Omega_M[1+f_2Q]$. It follows that the Friedmann equation has the same explicit form
\be
E^2(z)=\frac{H^2(z)}{H_0^2}=\Omega_\Lambda+\Omega_Mr+(1-\Omega_M-\Omega_\Lambda)r^{2n}\,,
\ee
although of course the redshift dependence is different. In this case the continuity equation can also be written in a way that elliminates $Q$,
\be
\frac{dr}{dz}=\frac{3r}{1+z}(1+w)\times\frac{\Omega_Mf_2+n(1-\Omega_M)f_1r^{2n-1}}{f_2[\Omega_M+2n(1-\Omega_M)r^{2n-1}]}\,.
\ee

%%%%%%%%%%%%%%%%%%%%%%%%%%%%%%%%%%%%%%%%%%%%%%%%%%%%%%%%%%%%%%%%%%%%%%%%%%
\subsection{The $\Omega_\Lambda=0$ matter case}
\label{casenolam}

We can now use our Pantheon and Hubble parameter data to constrain the scenario in which the recent acceleration of the universe is due to the nonlinear part of the Lagrangian rather than the usual cosmological constant. As seen above, in this case the parameter $Q$ can be elliminated so we have a two-dimensional parameter space $n$--$\Omega_M$ to constrain. As before the Hubble constant $H_0$ is analytically marginalized.

\begin{figure*}
\centering
\includegraphics[width=7.5cm]{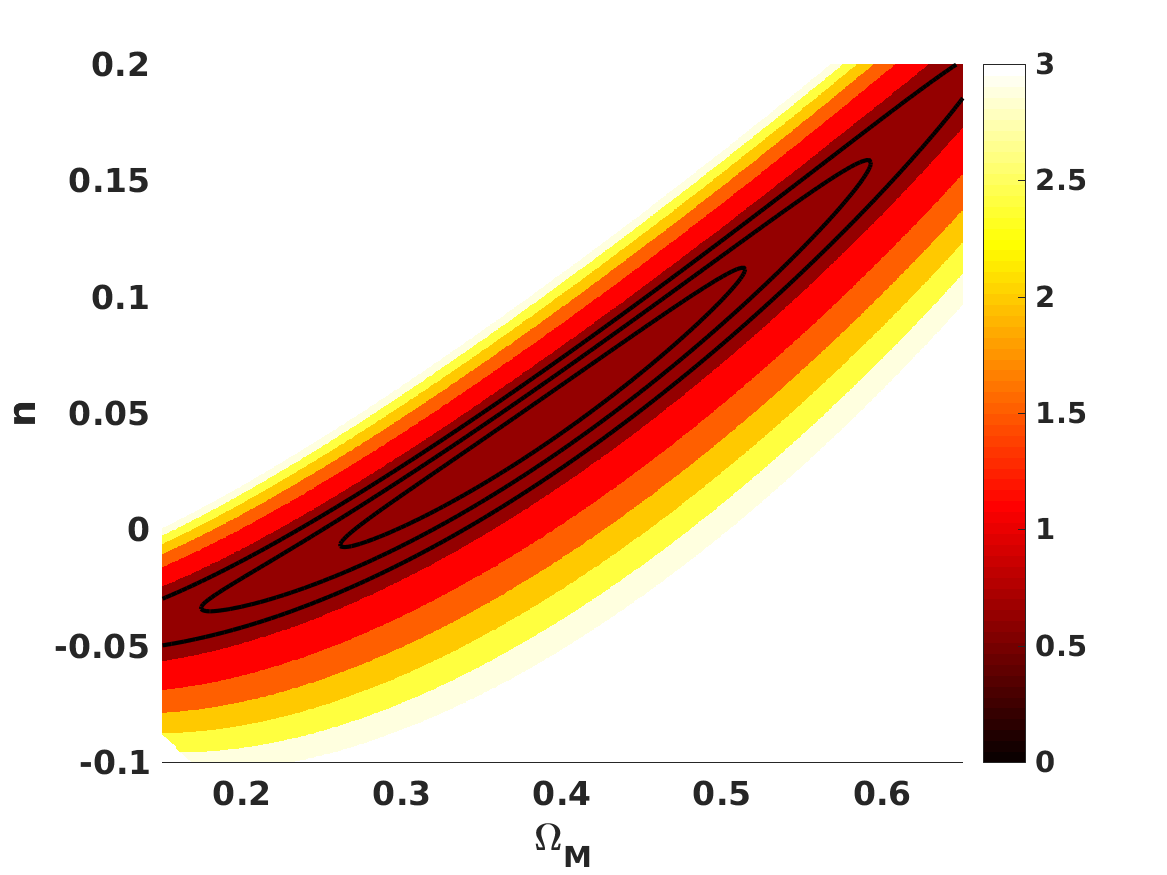}\\
\includegraphics[width=7.5cm]{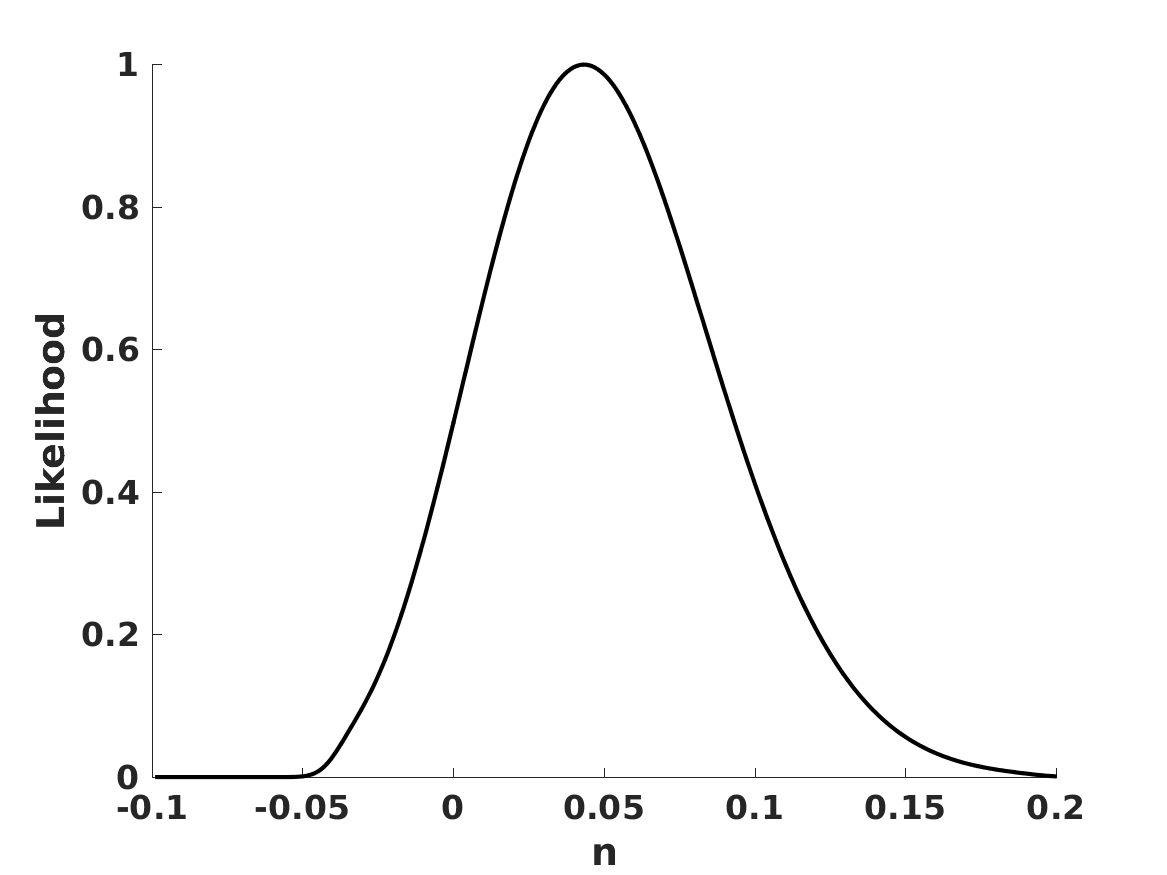}
\includegraphics[width=7.5cm]{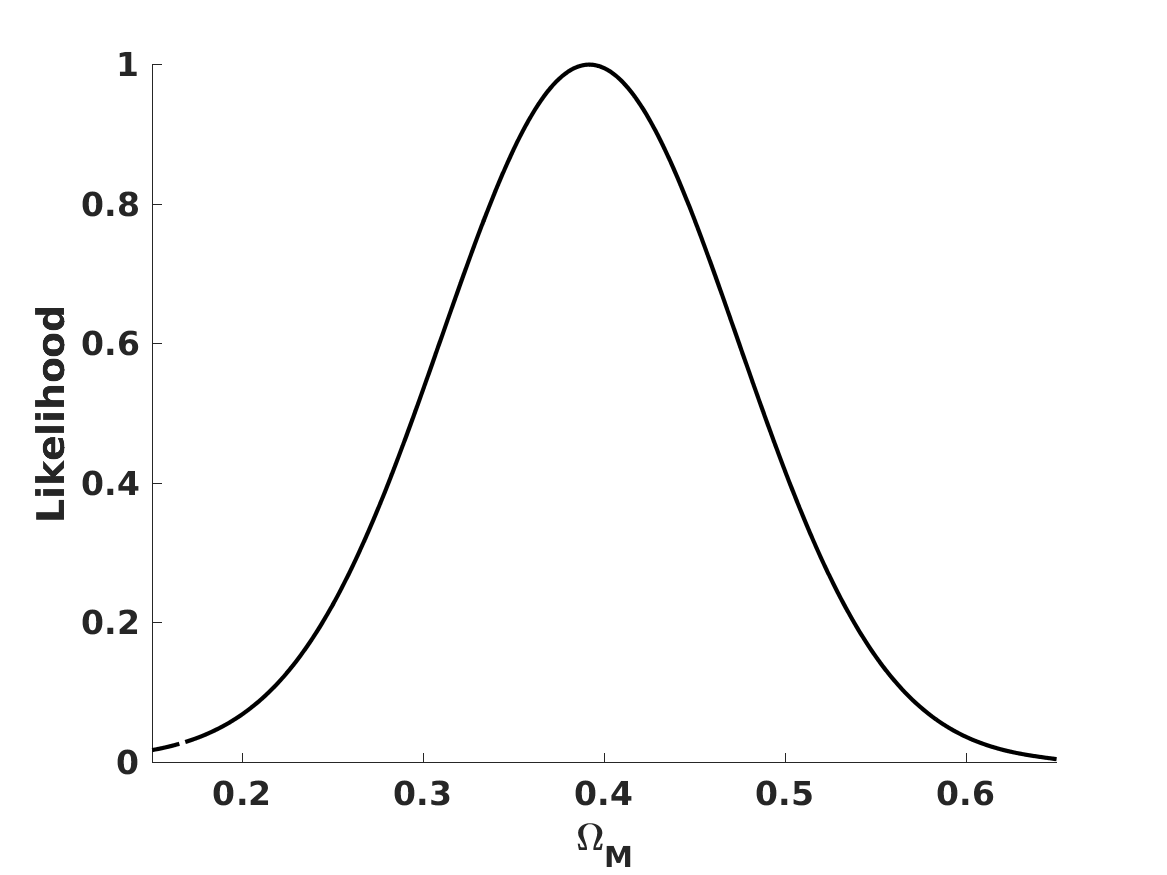}
\caption{Current constraints on the $n$--$\Omega_M$ parameter space for flat universes with $\Omega_\Lambda=0$. In the top panel the black solid curves show the one, two, and three sigma confidence levels in the two-dimensional plane, while the color map depicts the reduced chi-square. The bottom panels show the one-dimensional posterior likelihoods for both parameters.}
\label{figure10}
\end{figure*}

The results of this analysis are summarized in Fig. \ref{figure10} and also in the first row of Table \ref{table5}. As expected given the form of the Friedmann and continuity equations, there is a clear degeneracy between the two parameters. The best-fit values are about one standard deviation away from the canonical values $n=0$ and $\Omega_M\sim0.3$, and a non-zero $n\sim0.04$ and a slightly higher matter density are preferred. However, at the two sigma level the results are consistent with $\Lambda$CDM.

%%%%%%%%%%%%%%%%%%%%
%%%%%%%%%%%%%%%%%%%%

%%%%%%%%%%%%%%%%%%%%%%%%%%%%%%%%%%%%%%%%%%%%%%%%%%%%%%%%%%%%%%%%%%%%%%%%%%
\subsection{The $\Omega_\Lambda\neq0$ matter case}
\label{casewithlam}

In this case we have a three-dimensional parameter space to constrain, specifically $(\Omega_M,n,Q)$. It is also clear that there is a strong degeneracy between the model parameters $n$ and $Q$, and indeed we find that current low redshift data cannot constrain $Q$, and only constrains $n$ weakly. Nevertheless, the matter density can still be constrained.

Moreover, it is also clear that the constraints are somewhat dependent on the choice of priors. We explored two such choices. Firstly, if we allow $n$ to have any value from $n=0$ to $n=1$ and even slightly negative values, as we did already in the previous subsection, it is clear that $Q$ must be very small (in absolute value); this is because otherwise a term with a redshift dependence stronger than the matter one, $(1+z)^3$ would become dominant at fairly low redshifts. Secondly, if we exclude this scenario by imposing $n<le0.5$ then larger values of $|Q|$ are allowed. For these two scenarios we chose the uniform priors $|Q|\le0.05$ and $|Q|\le0.50$, respectively.

\begin{figure*}
\centering
\includegraphics[width=7.5cm]{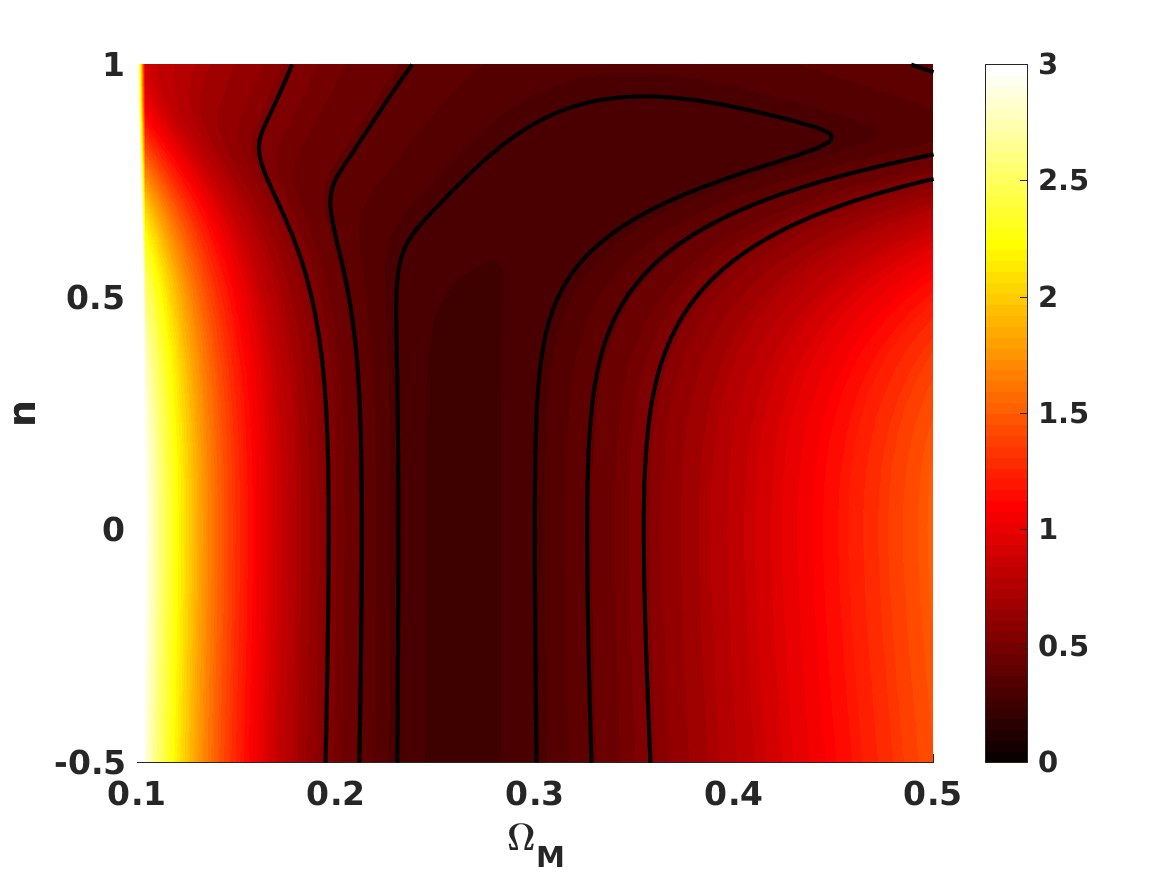}
\includegraphics[width=7.5cm]{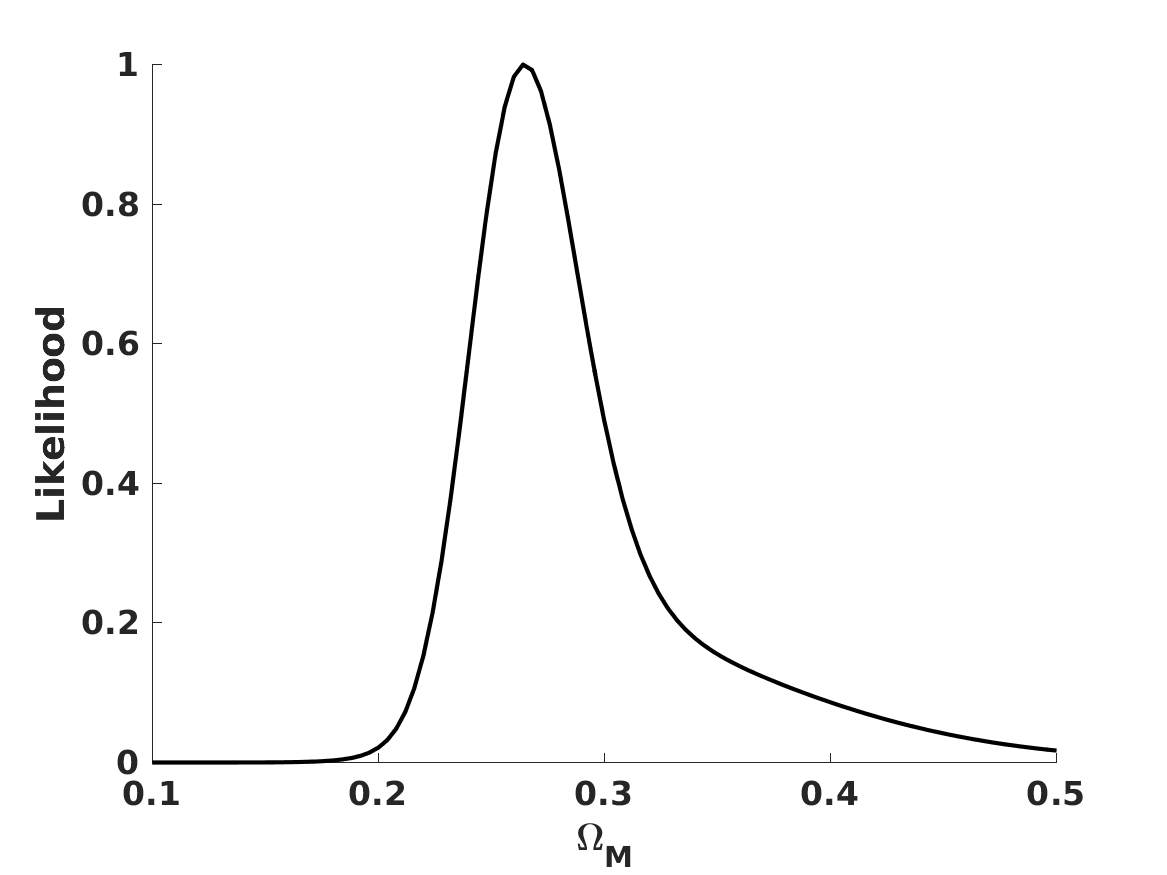}
\includegraphics[width=7.5cm]{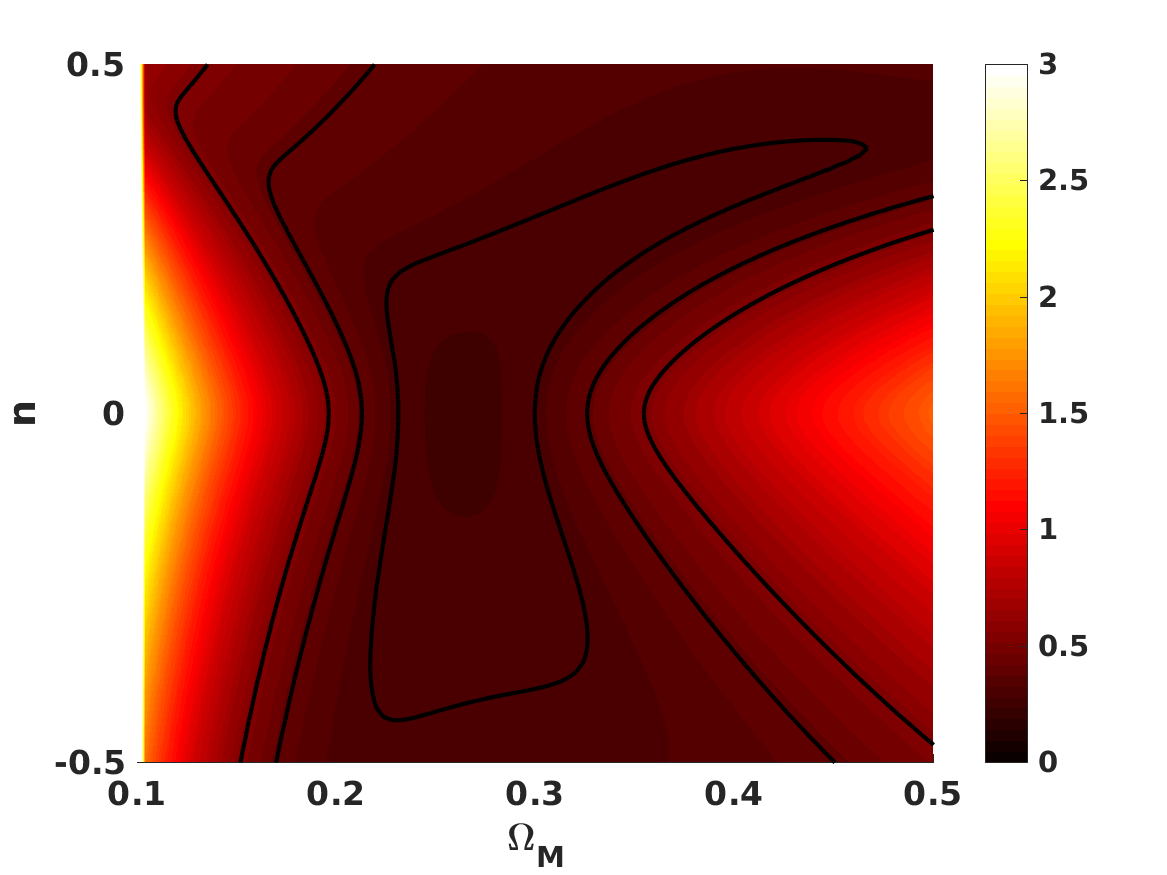}
\includegraphics[width=7.5cm]{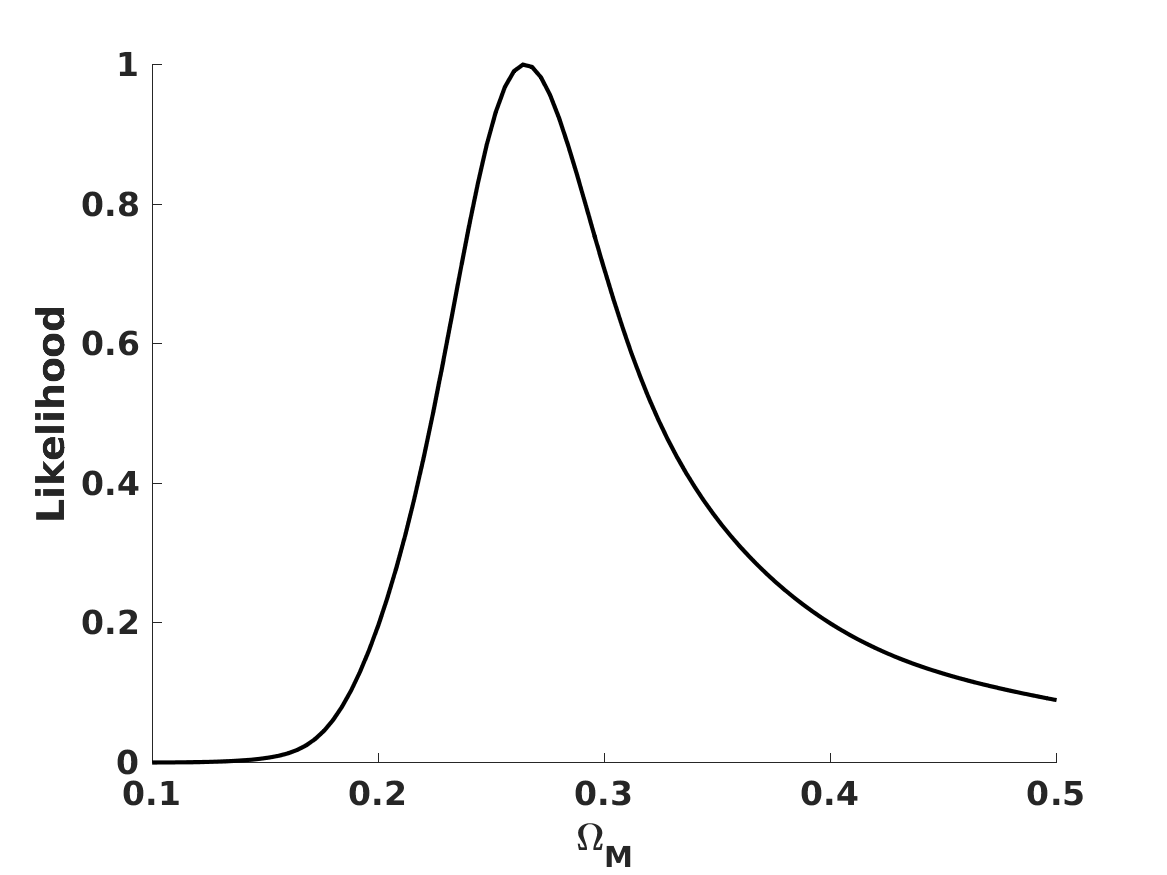}
\caption{Current constraints on the $n$--$\Omega_M$ parameter space for flat universes with $\Omega_\Lambda\neq0$. In the left panels the black solid curves show the one, two, and three sigma confidence levels in the two-dimensional plane, while the color map depict the reduced chi-square. The right panels show the one-dimensional posterior likelihoods for $\Omega_M$. The top and bottom panels correspond to the two choices of priors discussed in the text.}
\label{figure11}
\end{figure*}

The results  are summarized in Fig. \ref{figure11} and also in the second and third rows of Table \ref{table5}. With the first choice of priors, corresponding to small values of  $|Q|$, we can still constrain $n$ at the one-sigma level (with a large value preferred), but at the two sigma level it is unconstrained. On the other hand, allowing for larger $|Q|$ (our second choice of priors) $n$ becomes unconstrained. In either case, the posterior likelihood on $\Omega_M$ is only moderately affected by these choices and the best-fit value is actually the same in both cases, although the one-sigma uncertainties are significantly larger in the second case. This best-fit value is also slightly lower than the best-fit values for specific values of $n$, previously discussed in Section \ref{cnstr}, but still consistent with it given the larger uncertainties.

%%%%%%%%%%%%%%%%%%%%%%%%%%%%%%%%%%%%%%%%%%%%%%%%%%%%%%%%%%%%%%%%%%%%%%%%%%%%%%
\begin{table}
\caption{One sigma posterior likelihoods on the power $n$, the matter density $\Omega_M$ and the constant equation of state $w$ (when applicable) for various flat models containing matter, with or without a cosmological constant, and a nonlinear matter Lagrangian. The specific assumptions for each case are described in the main text. The constraints come from the combination of the Pantheon supernova data and Hubble parameter measurements.}
\label{table5}
\centering
\begin{tabular}{| c | c | c | c |}
\hline
Model assumptions & $n$ & $\Omega_M$ & $w$ \\
\hline
$\Omega_\Lambda=0$, Matter & $0.04\pm0.04$ & $0.39\pm0.08$ & N/A \\
\hline
$\Omega_\Lambda\neq0$, $|Q|\le0.05$ & $0.82^{+0.11}_{-0.26}$ & $0.26^{+0.03}_{-0.02}$ & N/A \\
$\Omega_\Lambda\neq0$, $|Q|\le0.50$ & Unconstrained & $0.26^{+0.05}_{-0.03}$ & N/A \\
\hline
$\Omega_\Lambda=0$, $w=const.$ & Unconstrained & $0.56^{+0.10}_{-0.27}$ & $-0.16^{+0.06}_{-0.05}$ \\
\hline
\end{tabular}
\end{table}
%%%%%%%%%%%%%%%%%%%%%%%%%%%%%%%%%%%%%%%%%%%%%%%%%%%%%%%%%%%%%%%%%%%%%%%%%%%%%%

%%%%%%%%%%%%%%%%%%%%
%%%%%%%%%%%%%%%%%%%%

%%%%%%%%%%%%%%%%%%%%%%%%%%%%%%%%%%%%%%%%%%%%%%%%%%%%%%%%%%%%%%%%%%%%%%%%%%
\subsection{The $\Omega_\Lambda=0$, $w=const$ case}
\label{casegenw}

Finally we return to the $\Omega_\Lambda=0$ case while allowing for a constant equation of state, $w=const$;  the matter case corresponds to $w=0$ In this case we also have a three-dimensional parameter space to constrain, specifically $(\Omega_M,n,w)$. It is again clear that we have a strong degeneracy between $n$ and $w$, which are both part of the two functions $f_1$ and $f_2$. It is clear that the region of parameter space, which is of interest for $\Omega_\Lambda=0$, is that around $n\sim0$ and $w\sim0$.

\begin{figure*}
\centering
\includegraphics[width=7.5cm]{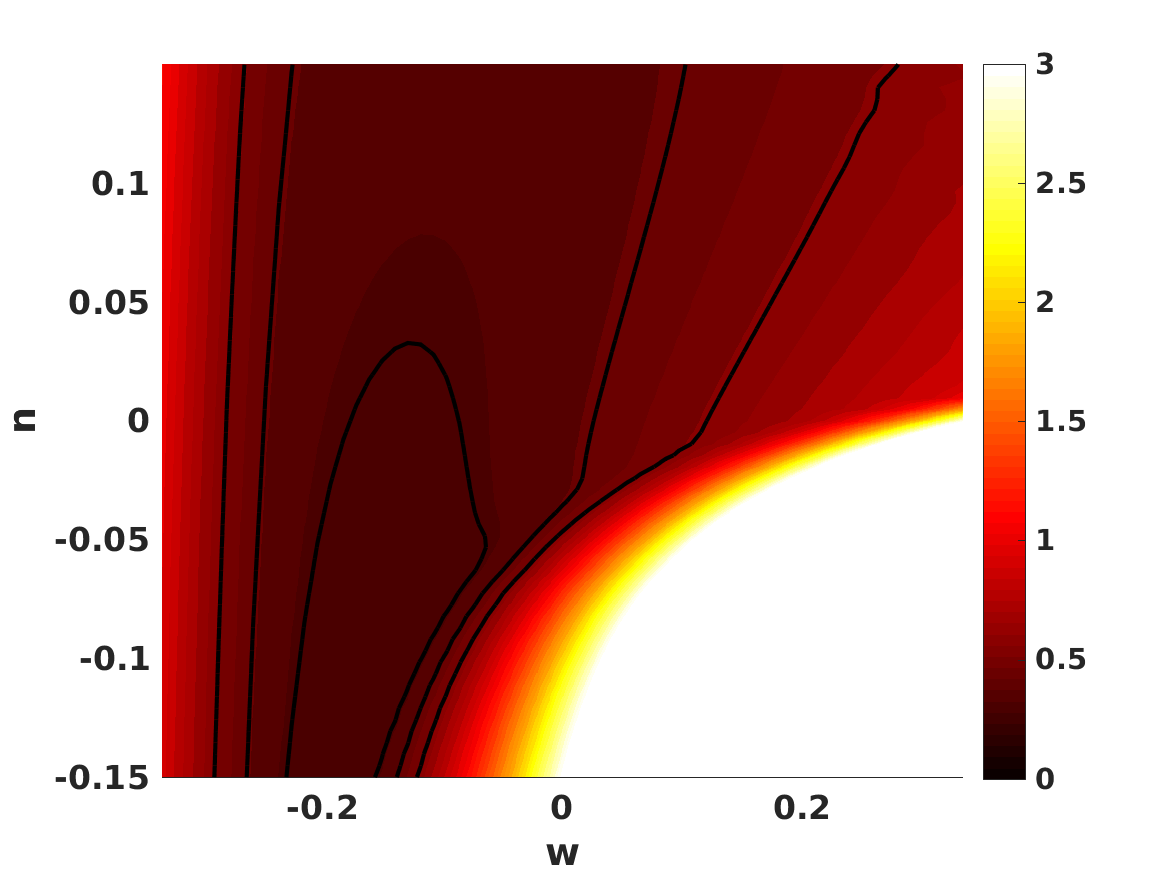}
\includegraphics[width=7.5cm]{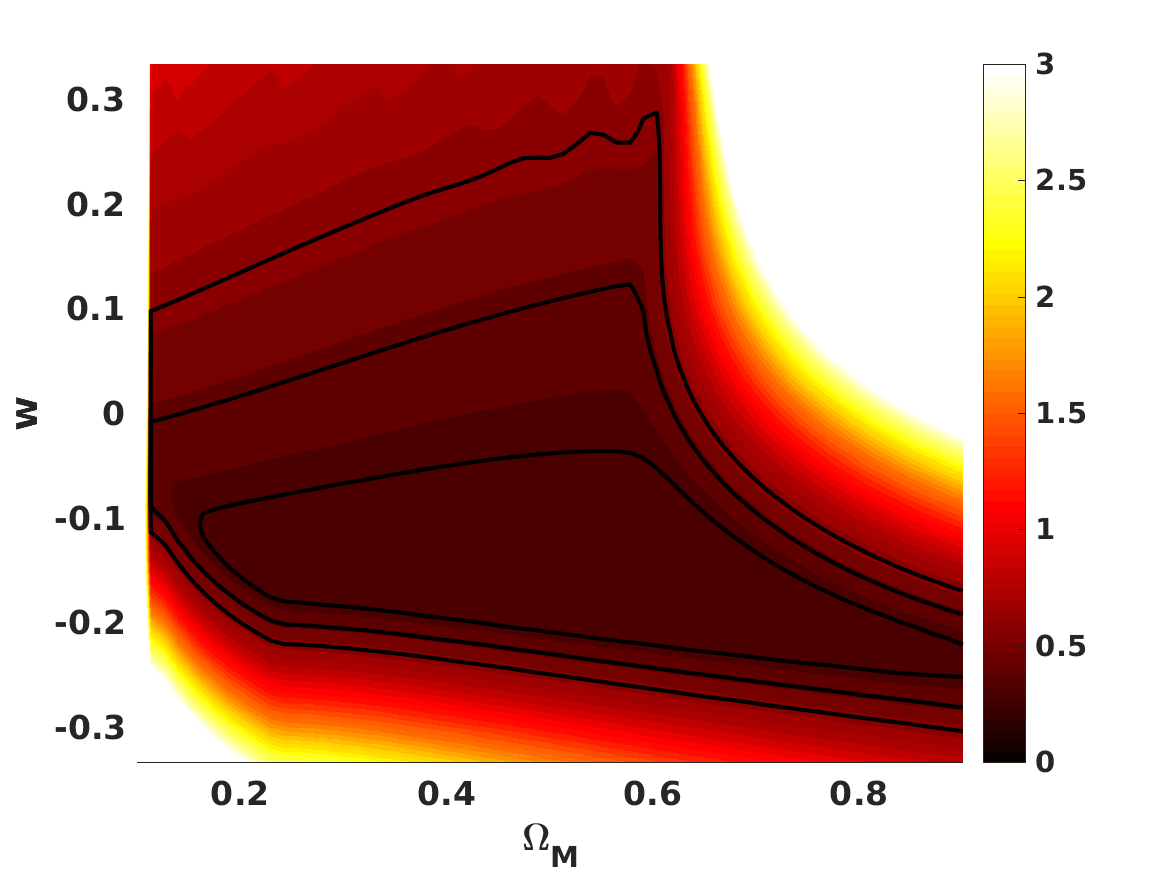}
\includegraphics[width=7.5cm]{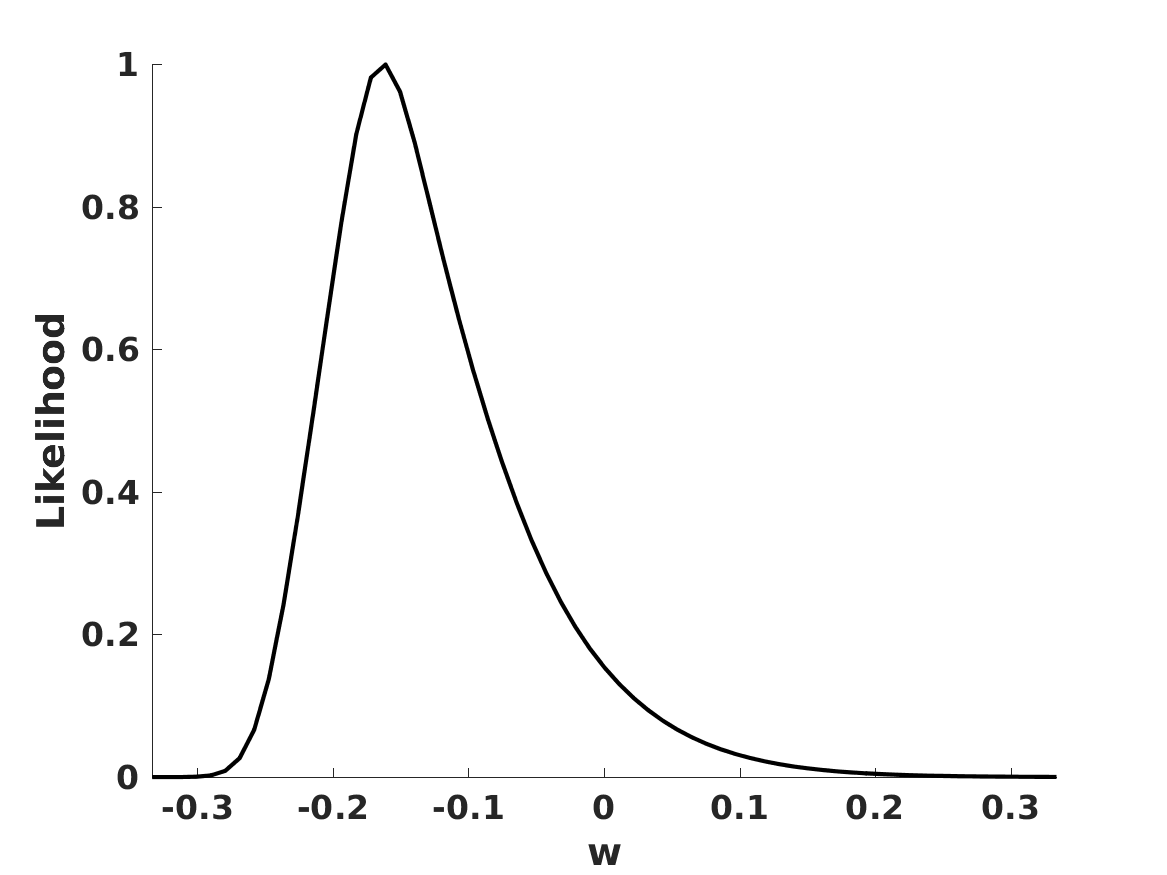}
\includegraphics[width=7.5cm]{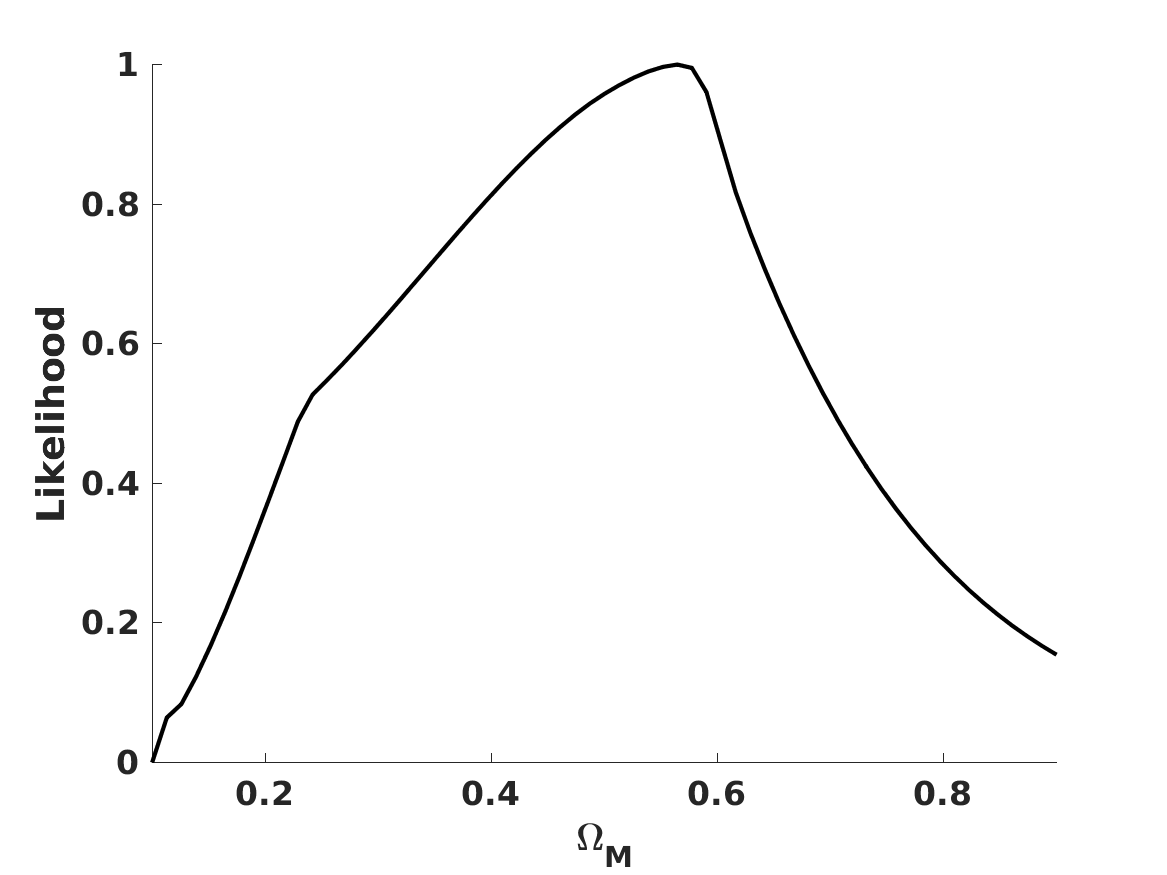}
\caption{Current constraints on the $n$--$\Omega_M$--$w$ parameter space for flat universes with $\Omega_\Lambda=0$. In the top panels the black solid curves show the one, two, and three sigma confidence levels in the two-dimensional plane, while the color maps depict the reduced chi-square. The bottom panels show the one-dimensional posterior likelihoods for the matter density and its equation of state (which is assumed to be constant).}
\label{figure12}
\end{figure*}

Our results are summarized in Fig. \ref{figure12} and also in the second and third rows of Table \ref{table5}. We find that $n$ is again unconstrained, while the matter density and its equation of state can be reasonably constrained. In the former case, however, the one-sigma uncertainties are significantly larger than in the $w=0$ cases studied in the rest of the article. Again these is a preference for a higher matter density (as well as slightly negative equation of state), although given the larger error bars this is not statistically significant.

%%%%%%%%%%%%%%%%%%%%
%%%%%%%%%%%%%%%%%%%%

%%%%%%%%%%%%%%%%%%%%%%%%%%%%%%%%%%%%%%%%%%%%%%%%%%%%%%%%%%%%%%%
\section{\label{concl}Conclusions}

We have presented a quantitative study of the low redshift cosmological consequences of energy-momentum-powered gravity models, in which the matter part of Einstein's equations is modified by the addition of a term proportional to some power of the energy-momentum tensor, in addition to the canonical linear term. These models can be  phenomenologically considered as extensions of general relativity with a nonlinear matter Lagrangian. In this sense they are interesting since they are somewhat different from the usual dynamical dark energy or modified gravity models, which either introduce extra dynamical degrees of freedom to the Lagrangian (typically scalar fields) or gravitational part of the Lagrangian. In our study we pursued two different approaches, treating these models either as extensions of the canonical $\Lambda$CDM scenario or as more direct alternatives to  this scenario (in the sense of being capable of producing late-time acceleration without a true cosmological constant).

In the first approach, these models include additional parameters that are constrained to be small, so that the standard $\Lambda$CDM model is recovered in the appropriate limits. In the early part of our work we focused on three specific examples of these models (characterized by three different powers of the nonlinear matter Lagrangian term, given by the model parameter $n$), for which  there are solutions that are analytic solutions and lead to models with a clear physical interpretation, and we constrained the corresponding parameters using recent low redshift background cosmology data. We also briefly discussed how next generation ground and space astrophysical facilities will improve on these constraints. In the latter part of the work we also studied the broader parameter space where the power of the nonlinear term, $n$, is itself a free parameter.

As a second approach we analyzed the previous suggestion that these models might---at least at a penomenological level---account for the recent acceleration of the universe without invoking a cosmological constant (i.e., with only a matter component). We showed that this is in principle a possibility and constrained the corresponding value of $n$, which needs to be no larger (in absolute value) than a few percent. Interestingly, in this case there is a positive correlation between $n$ and the matter density: a slightly positive value of $n$ requires a value of $\Omega_M$, which is slightly larger than the canonical value. For this case we also relaxed the assumption that the single component is matter, allowing it to have a constant equation of state; in this case a slightly negative equation of state is preferred, while the preferred value of $\Omega_M$ is also larger. In all cases, the preferred values differ from the canonical $\Lambda$CDM by no more than two standard deviations.

We note that our work was only concerned with low redshift background cosmology constraints. Two interesting topics that we have not addressed, but leave for future work, are the inclusion of high redshift constraints, in particular from the cosmic microwave background, and, more generally, the clustering properties of these models. The field equations of the model differ from those of general relativity only in the presence of matter sources, so we might also expect differences at this level. These may lead to constraints on this class of models that are more stringent than those we presented here. In any case, the extent to which a nonlinear matter Lagrangian may contribute to the dynamics of the recent universe is certainly an interesting question for further exploration.

%%%%%%%%%%%%%%%%%%%%%%%%%%%%%%%%%%%%%%%%%%%%%%%%%%%%%%%%%%%

\begin{acknowledgements}
This work was financed by FEDER---Fundo Europeu de Desenvolvimento Regional funds through the COMPETE 2020---Operacional Programme for Competitiveness and Internationalisation (POCI), and by Portuguese funds through FCT---Funda\c c\~ao para a Ci\^encia e a Tecnologia in the framework of the project POCI-01-0145-FEDER-028987. Some of the work described herein was done at or following AstroCamp 2018.
\end{acknowledgements}

\bibliographystyle{aa} % style aa.bst
\bibliography{square} % your references Yourfile.bib

\begin{thebibliography}{22}
\expandafter\ifx\csname natexlab\endcsname\relax\def\natexlab#1{#1}\fi

\bibitem[{Abbott {et~al.}(2018)}]{DESY1}
Abbott, T. M.~C. {et~al.} 2018, Phys. Rev., D98, 043526

\bibitem[{Aghanim {et~al.}(2018)}]{Planck}
Aghanim, N. {et~al.} 2018 [\eprint[arXiv]{1807.06209}]

\bibitem[{Akarsu {et~al.}(2018)Akarsu, Katirci, \& Kumar}]{Akarsu}
Akarsu, O., Katirci, N., \& Kumar, S. 2018, Phys. Rev., D97, 024011

\bibitem[{Anagnostopoulos \& Basilakos(2018)}]{Homarg}
Anagnostopoulos, F.~K. \& Basilakos, S. 2018, Phys. Rev., D97, 063503

\bibitem[{Board \& Barrow(2017)}]{Board}
Board, C. V.~R. \& Barrow, J.~D. 2017, Phys. Rev., D96, 123517, [Erratum: Phys.
  Rev.D98,no.12,129902(2018)]

\bibitem[{Concas {et~al.}(2017)Concas, Pozzetti, Moresco, \& Cimatti}]{Concas}
Concas, A., Pozzetti, L., Moresco, M., \& Cimatti, A. 2017, Mon. Not. Roy.
  Astron. Soc., 468, 1747

\bibitem[{Farooq {et~al.}(2017)Farooq, Madiyar, Crandall, \& Ratra}]{Farooq}
Farooq, O., Madiyar, F.~R., Crandall, S., \& Ratra, B. 2017, Astrophys. J.,
  835, 26

\bibitem[{Freese \& Lewis(2002)}]{Cardassian}
Freese, K. \& Lewis, M. 2002, Phys. Lett., B540, 1

\bibitem[{Jimenez \& Loeb(2002)}]{Jimenez}
Jimenez, R. \& Loeb, A. 2002, Astrophys. J., 573, 37

\bibitem[{Katırcı \& Kavuk(2014)}]{Early}
Katırcı, N. \& Kavuk, M. 2014, Eur. Phys. J. Plus, 129, 163

\bibitem[{Liske {et~al.}(2008)}]{Liske}
Liske, J. {et~al.} 2008, Mon. Not. Roy. Astron. Soc., 386, 1192

\bibitem[{Liske {et~al.}(2014)}]{HIRES}
Liske, J. {et~al.} 2014, {Top Level Requirements For ELT-HIRES}, Tech. rep.,
  Document ESO 204697 Version 1

\bibitem[{{Liu} {et~al.}(2016){Liu}, {Lu}, {Xie}, {Chen}, \& {Zhao}}]{LiuLu}
{Liu}, G.~C., {Lu}, Y.~J., {Xie}, L.~Z., {Chen}, X.~L., \& {Zhao}, Y.~H. 2016,
  Astron. Astrophys., 585, A52

\bibitem[{Lopez-Corredoira \& Vazdekis(2018)}]{Corredoira}
Lopez-Corredoira, M. \& Vazdekis, A. 2018, Astron. Astrophys., 614, A127

\bibitem[{Lopez-Corredoira {et~al.}(2017)Lopez-Corredoira, Vazdekis, Gutierrez,
  \& Castro-Rodriguez}]{Vazdekis}
Lopez-Corredoira, M., Vazdekis, A., Gutierrez, C.~M., \& Castro-Rodriguez, N.
  2017, Astron. Astrophys., 600, A91

\bibitem[{Poplawski(2006)}]{Poplawski}
Poplawski, N.~J. 2006 [\eprint[arXiv]{gr-qc/0608031}]

\bibitem[{Riess {et~al.}(2018)}]{Riess}
Riess, A.~G. {et~al.} 2018, Astrophys. J., 853, 126

\bibitem[{Roshan \& Shojai(2016)}]{Roshan}
Roshan, M. \& Shojai, F. 2016, Phys. Rev., D94, 044002

\bibitem[{{Sandage}(1962)}]{Sandage}
{Sandage}, A. 1962, Ap. J., 136, 319

\bibitem[{Spergel {et~al.}(2015)}]{WFIRST}
Spergel, D. {et~al.} 2015 [\eprint[arXiv]{1503.03757}]

\bibitem[{Suzuki {et~al.}(2012)}]{Suzuki}
Suzuki, N. {et~al.} 2012, Astrophys. J., 746, 85

\bibitem[{Tutusaus {et~al.}(2016)Tutusaus, Lamine, Blanchard, Dupays, Rousset,
  \& Zolnierowski}]{Tutusaus}
Tutusaus, I., Lamine, B., Blanchard, A., {et~al.} 2016, Phys. Rev., D94, 123515

\end{thebibliography}

\begin{appendix}
\section{Comparison with a dynamical dark energy model}

As mentioned in the introduction, these energy-momentum-powered gravity models are somewhat different from the usual dynamical dark energy and modified gravity models. In the former class of models further dynamical degrees of freedom are added to the Lagrangian (often in the form of scalar fields), while in the latter the gravitational part of the Lagrangian is changed. We provide a simple illustrative comparison with a dynamical dark energy model. We again use the compressed data from the Pantheon compilation in \citet{Riess} and the compilation of Hubble parameter measurements from \citet{Farooq}, and analytically marginalize the Hubble constant.

\begin{figure*}
\centering
\includegraphics[width=7.5cm]{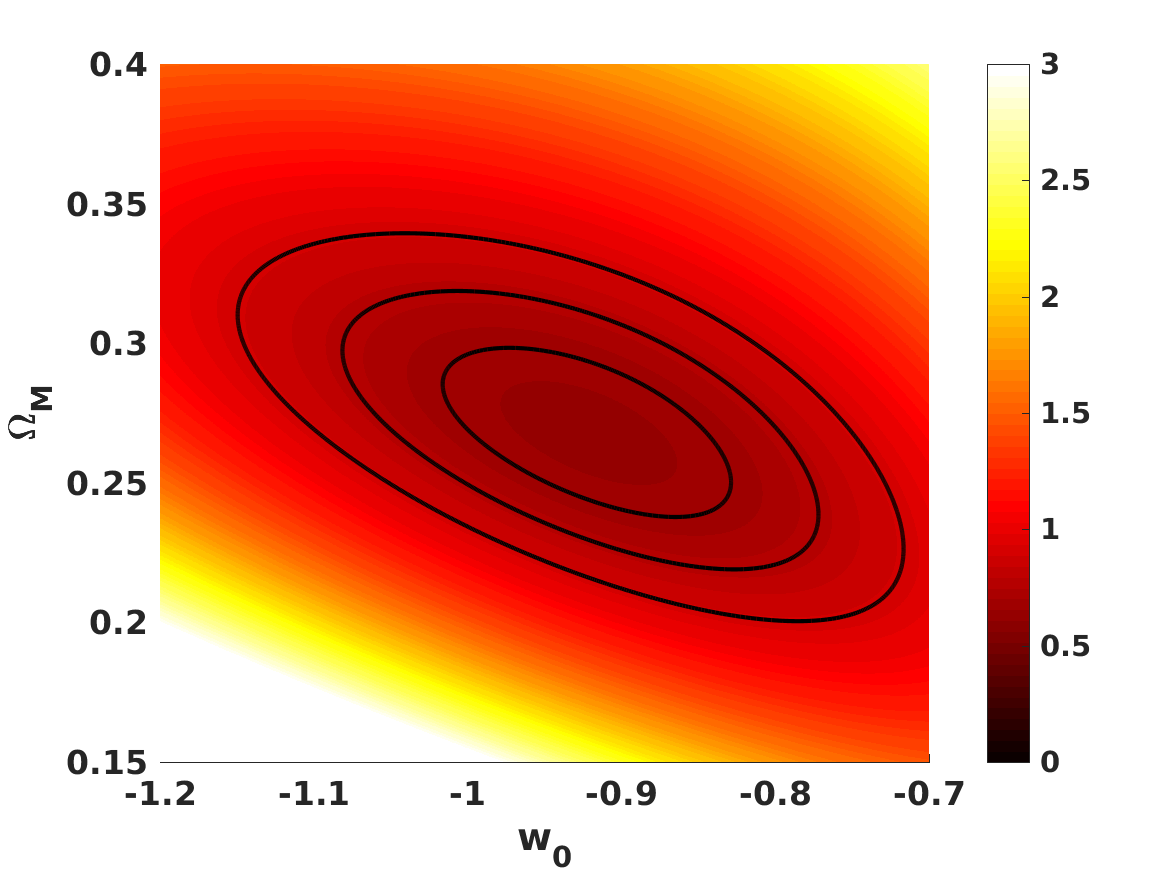}
\includegraphics[width=7.5cm]{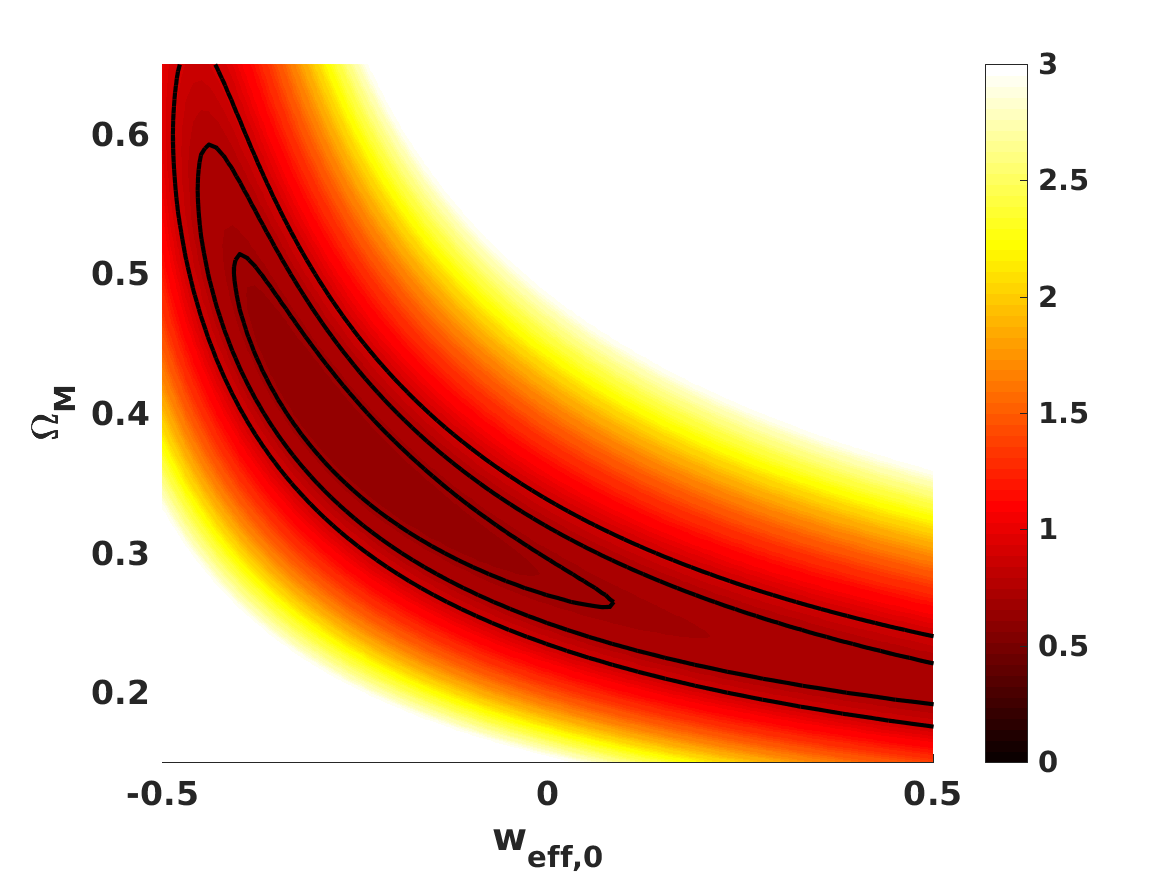}
\includegraphics[width=7.5cm]{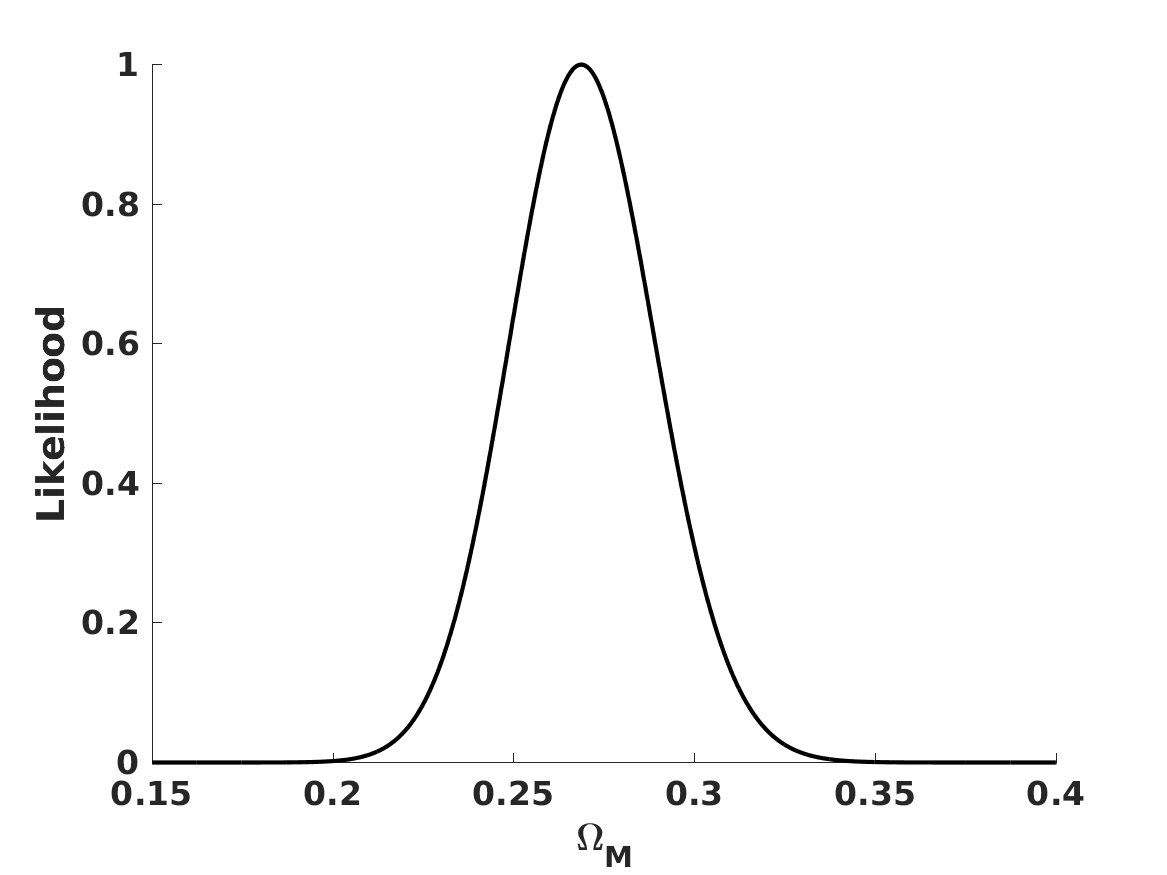}
\includegraphics[width=7.5cm]{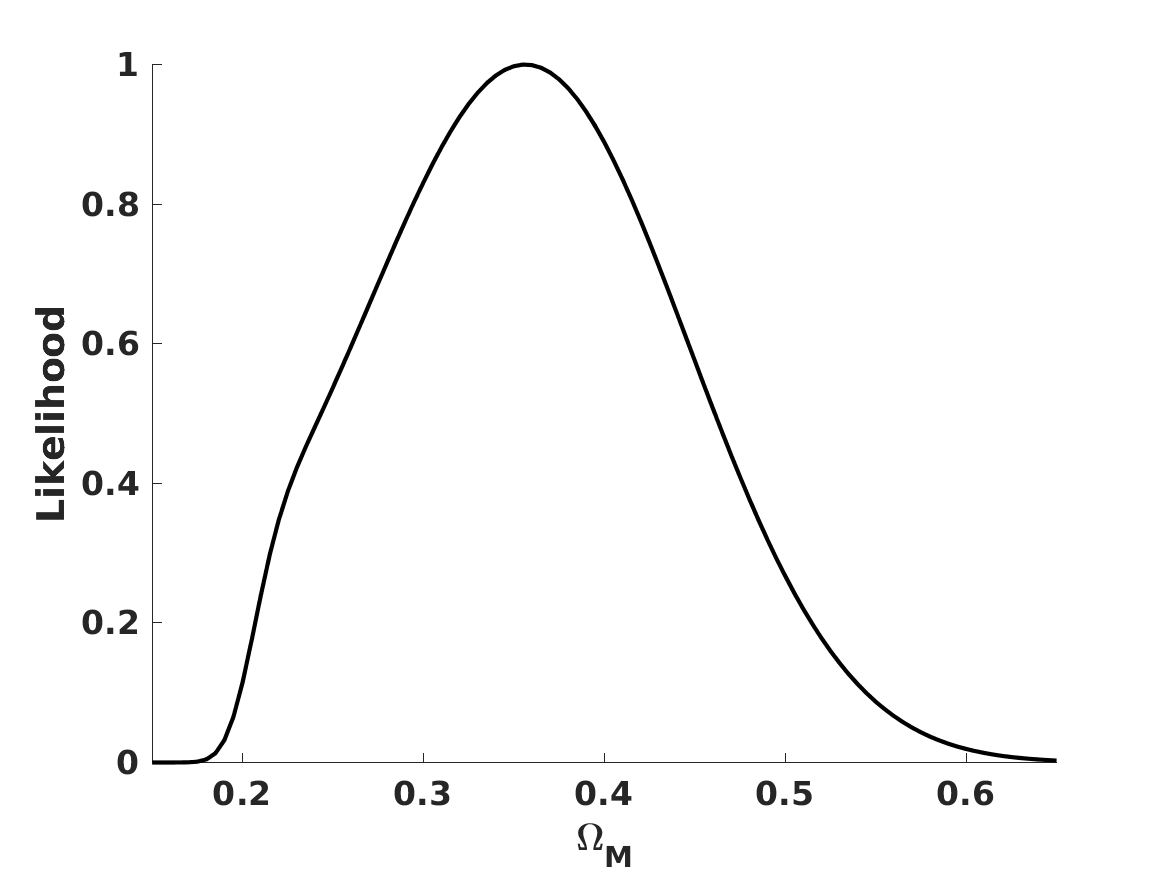}
\includegraphics[width=7.5cm]{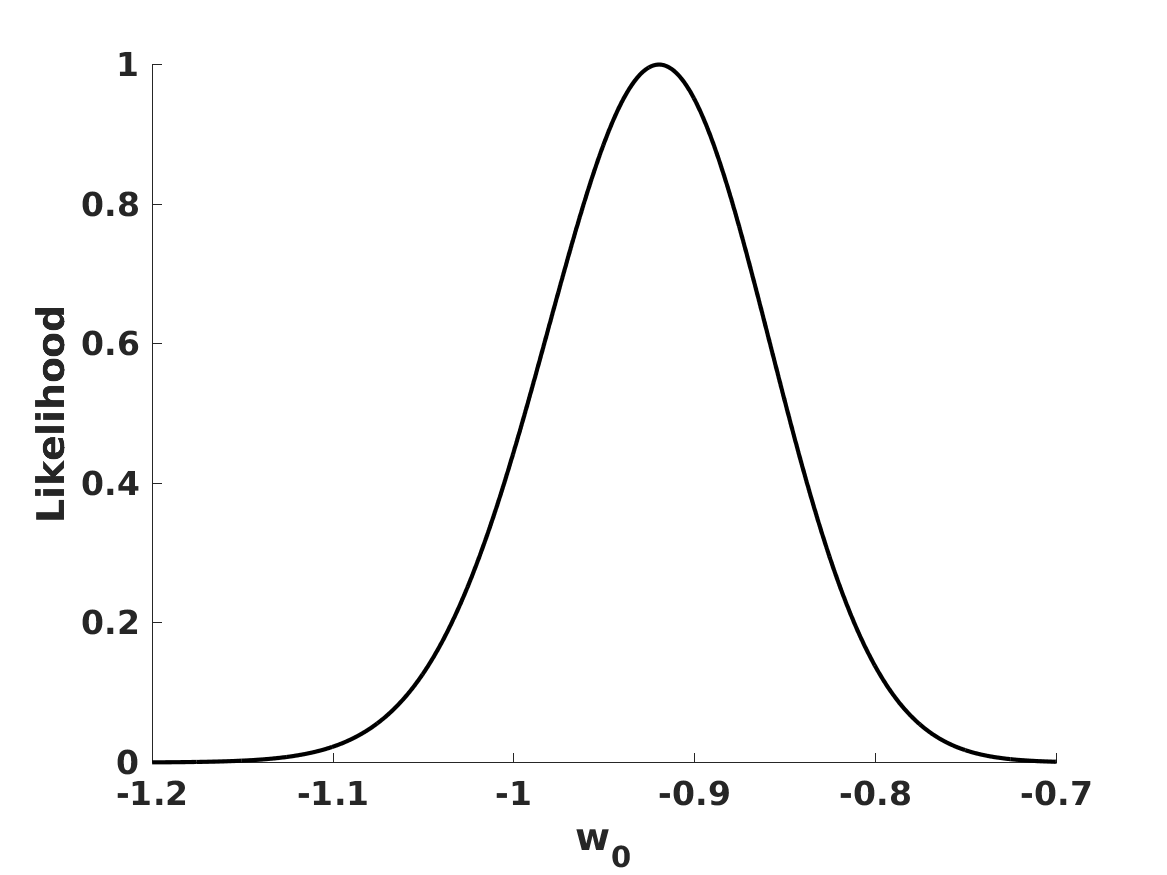}
\includegraphics[width=7.5cm]{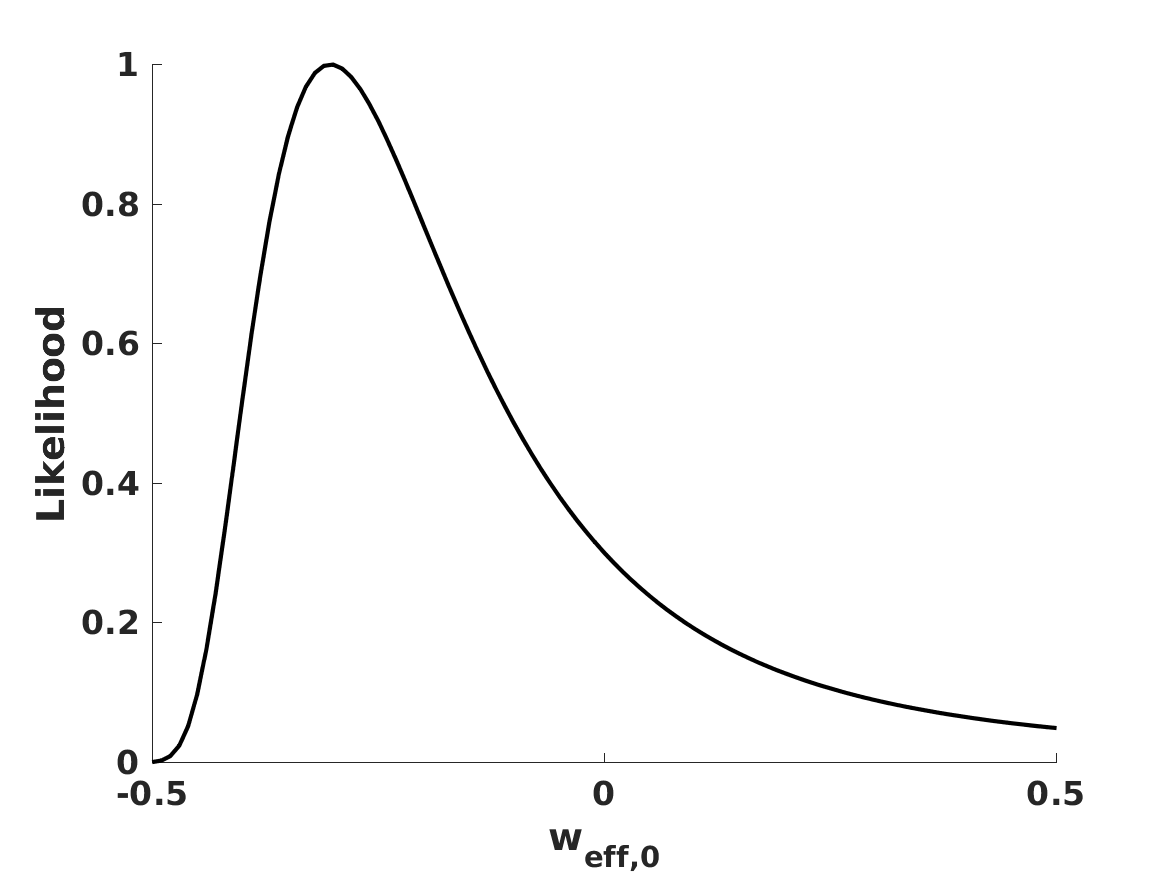}
\caption{Comparing constraints in a dynamical dark energy model with a constant equation of state (left-hand side panels) and in an effective description of an energy-momentum-powered gravity model as a dark energy model with an effective equation of state (right-hand side panels). In the top panels the black solid curves show the one, two, and three sigma confidence levels in the two-dimensional plane, while the color maps depict the reduced chi-square. The middle and bottom panels show the one-dimensional posterior likelihoods for the matter density and the relevant equation of state parameters defined in the text.}
\label{figure13}
\end{figure*}

For a canonical dynamical dark energy model, assumed to be flat and with a constant dark energy equation of state, $w_0$, the Friedmann equation reads
\be
\frac{H^2(z)}{H_0^2}=\Omega_M(1+z)^3+(1-\Omega_M)(1+z)^{3(1+w_0)}\,,
\ee
and the continuity equation for the dark energy is the standard\be
\frac{dr}{dz}=\frac{3r}{1+z}(1+w_0)\,.
\ee
The constraints on this model from our datasets are depicted on the left-hand side of Figure \ref{figure13}. The one sigma posterior likelihoods for the two model parameters are written as
\be
\Omega_M=0.27\pm0.02
\ee
\be
w_0=-0.92\pm0.06\,.
\ee
At the two sigma level this is consistent with $\Lambda$CDM, and the acceleration is due to a component whose equation of state is fairly similar to that of a cosmological constant.

Now consider the energy-momentum-powered gravity models studied in Section \ref{casenolam}, also assumed to be flat. In this case the only component of the universe is matter, but the nonlinear Lagrangian leads to additional terms in the Friedmann equation, which now reads
\be
E^2(z)=\frac{H^2(z)}{H_0^2}=\Omega_Mr+(1-\Omega_M)r^{2n}\,,
\ee
and the continuity equation now has the form
\be
\frac{dr}{dz}=\frac{3r}{1+z}\times\frac{1+nQr^{2n-1}}{1+(2n-1)nQr^{2n-1}}\,.
\ee
We could rewrite this equation in a form akin to that of a dynamical dark energy model, defining an effective equation of state
\be
\frac{dr}{dz}=\frac{3r}{1+z}(1+w_{eff}(z))\,;
\ee
we note that this effective equation of state is redshift-dependent. In particular, at redshift $z=0$ we have
\be
w_{eff,0}=\frac{2n(1-n)Q}{1+(2n-1)nQ}
\ee
and the flatness condition also implies
\be
Q\left(n-\frac{1}{2}\right)=\frac{1-\Omega_M}{\Omega_M}\,.
\ee
We see that the effective equation of state is indeed be matter-like ($w_{eff,0}=0$) if $Q=0$ or $n=0$, but is different otherwise. We can repeat the analysis of Section \ref{casenolam}, using the above relations to constrain the parameter space $(\Omega_M,w_{eff,0})$ instead of $(\Omega_M,n)$. The results are similar although not identical, since a uniform prior on $n$ (assumed in Section \ref{casenolam}) is not equivalent to a uniform prior on $w_{eff,0}$ (assumed here) since the relation between the two parameters is nonlinear. The constraints on the $(\Omega_M,n)$ parameter space are shown in Figure \ref{figure10}; those on the $(\Omega_M,w_{eff,0})$ parameter space are in the right-hand side of Figure \ref{figure13}. In this case the one-sigma posterior likelihoods on the two model parameters are
written as\be
\Omega_M=0.36^{+0.08}_{-0.10}
\ee
\be
w_{eff,0}=-0.30^{+0.16}_{-0.09}\,.
\ee

This provides a simple but hopefully illustrative comparison between the two models. As expected the constraints are stronger for the ``standard'' dynamical dark energy model. As already discussed in the main text, we see that energy-momentum-powered models can only reproduce the recent acceleration of the universe without a true cosmological constant by having a matter density that is higher than that preferred by the $\Lambda$CDM scenario. On the other hand the present-day value of the effective equation of state is somewhat different from the matter one ($w_m=0$), but compatible with that obtained (under different assumptions) in Section \ref{casegenw}.

%%%%%%%%%%%%%%%%%%%%
%%%%%%%%%%%%%%%%%%%%

\end{appendix}

\end{document}